\documentclass[%
 aip,
 jmp,%
 amsmath,amssymb,
 reprint,%
]{revtex4-1}

\usepackage{graphicx}
\usepackage{dcolumn}
\usepackage{bm}

\usepackage{subfig} 
\usepackage{url} 
\usepackage{xcolor} 
\definecolor{newcolor}{rgb}{.8,.349,.1} 
\usepackage{booktabs,multirow} 
\usepackage{lineno}
\usepackage[normalem]{ulem} 
\usepackage{gensymb} 

\begin{document}

\title[Journal of Applied Physics]{Explosively driven Richtmyer--Meshkov instability jet suppression and enhancement via coupling machine learning and additive manufacturing}

\affiliation{Lawrence Livermore National Laboratory, Livermore, CA 94550, USA}%

\author{Dane M. Sterbentz}
\email{sterbentz2@llnl.gov}
\affiliation{Lawrence Livermore National Laboratory, Livermore, CA 94550, USA}%
\affiliation{These authors contributed equally to this work.}%

\author{Dylan J. Kline}%
\email{kline11@llnl.gov}
\affiliation{Lawrence Livermore National Laboratory, Livermore, CA 94550, USA}%
\affiliation{These authors contributed equally to this work.}%

\author{Daniel A. White}%
\affiliation{Lawrence Livermore National Laboratory, Livermore, CA 94550, USA}%

\author{Charles F. Jekel}
\affiliation{Lawrence Livermore National Laboratory, Livermore, CA 94550, USA}%

\author{Michael P. Hennessey}%
\affiliation{Lawrence Livermore National Laboratory, Livermore, CA 94550, USA}%

\author{David K. Amondson}%
\affiliation{Lawrence Livermore National Laboratory, Livermore, CA 94550, USA}%

\author{Abigail J. Wilson}%
\affiliation{Lawrence Livermore National Laboratory, Livermore, CA 94550, USA}%

\author{Max J. Sevcik}%
\affiliation{Lawrence Livermore National Laboratory, Livermore, CA 94550, USA}%
\affiliation{Colorado School of Mines, Golden, CO, 80401, USA}%

\author{Matthew F. L. Villena}%
\affiliation{Lawrence Livermore National Laboratory, Livermore, CA 94550, USA}%

\author{Steve S. Lin}%
\affiliation{Lawrence Livermore National Laboratory, Livermore, CA 94550, USA}%

\author{Michael D. Grapes}%
\affiliation{Lawrence Livermore National Laboratory, Livermore, CA 94550, USA}%

\author{Kyle T. Sullivan}%
\affiliation{Lawrence Livermore National Laboratory, Livermore, CA 94550, USA}%

\author{Jonathan L. Belof}%
\email{belof1@llnl.gov}
\affiliation{Lawrence Livermore National Laboratory, Livermore, CA 94550, USA}%

\date{\today}

\begin{abstract}
The ability to control the behavior of fluid instabilities at material interfaces, such as the shock-driven Richtmyer--Meshkov instability, is a grand technological challenge with a broad number of applications ranging from inertial confinement fusion experiments to explosively driven shaped charges. In this work, we use a linear-geometry shaped charge as a means of studying methods for controlling material jetting that results from the Richtmyer--Meshkov instability. A shaped charge produces a high-velocity jet by focusing the energy from the detonation of high explosives. The interaction of the resulting detonation wave with a hollowed cavity lined with a thin metal layer produces the unstable jetting effect. By modifying characteristics of the detonation wave prior to striking the lined cavity, the kinetic energy of the jet can be enhanced or reduced. Modifying the geometry of the liner material can also be used to alter jetting properties. We apply optimization methods to investigate several design parameterizations for both enhancing or suppressing the shaped-charge jet. This is accomplished using 2D and 3D hydrodynamic simulations to investigate the design space that we consider. We also apply new additive manufacturing methods for producing the shaped-charge assemblies, which allow for experimental testing of complicated design geometries obtained through computational optimization. We present a direct comparison of our optimized designs with experimental results carried out at the High Explosives Application Facility at Lawrence Livermore National Laboratory.
\end{abstract}

\maketitle

\section{Introduction}
The Richtmyer--Meshkov instability (RMI) is a fluid instability that occurs in many high-pressure scenarios, both natural and technological, involving shock and ramp waves. At an interface between two materials of different densities, a shock wave passing through this interface deposits vorticity due to the baroclinic torque produced from misalignments in density and pressure gradients across a non-flat interface. This leads to the development of unstable fluid motion and RMI jetting at non-flat grooves or cavities in the interface. Depending on the application, it may be beneficial to either enhance RMI jetting or suppress this jetting. For instance, RMI is prevalent in inertial confinement fusion (ICF) experiments, where high-energy lasers may be used to compress and implode a fuel capsule using shock waves~\cite{Lindl1992,Hammer1999,Aglitskiy2010,Betti2016,Zylstra2022}. For ICF experiments, it is desirable to suppress material interface jetting during compression to reduce asymmetry in the implosion process and increase the energy yield of the experiments~\cite{Sterbentz2022}. On the other hand, explosively driven RMI jets or shaped charges, which focus energy from the detonation of high explosives (HE) to create a high-velocity metal jet for penetrating a given material, take advantage of the RMI jetting effect. This high-velocity jet is formed when a detonation wave strikes a metal liner that covers a cavity in the HE. Shaped charges are used in a number of industrial applications including rock blasting for mining applications~\cite{Persson2018}, emergency pipeline cutoff in the oil and gas industry~\cite{Ko2017,Cheng2019}, and booster stage separation for multistage rockets~\cite{Sun2020,Choi2022}. 

A number of recent works have demonstrated that modifying detonation wave properties and the interaction with the liner of various types of shaped charges can be used to alter and enhance jetting properties~\cite{Murphy1999,Wang2008,Zhang2017,Cheng2018,Ma2021,Xu2021,Wang2018,Liu2019a,Liu2019b,Pyka2020,Liu2020,Wu2023,Sterbentz2023,Kline2024,Hennessey2023}. A recent article by Kline and Hennessey et al.~\cite{Kline2024} demonstrated a method to augment jet velocities in explosively driven, additively manufactured linear shaped charges that featured an inert silicone buffer to modulate the shock front during the detonation event. However, the work performed by the authors showcases only a single design and method to augment the shaped charge jet. Another recent article by Sterbentz et al.~\cite{Sterbentz2023} describes a computational optimization methodology for increasing penetrative properties of a linear shaped charge jet using several design parameterizations. This article also describes some of the underlying physics for understanding how the detonation wave--liner  interaction affects jet mass and velocity and how detonation wave collisions affect jet formation. 

In this work, we expand upon the previous works and leverage the same experimental platform and methodologies in Kline and Hennessey et al.\ to investigate alternative techniques to enhance or suppress RMI jet formation in explosively driven linear shaped charges. To accomplish this, we use 2D and 3D hydrodynamic simulations to study a linear shaped charge baseline scenario and propose several novel design parameterizations to modify jet formation by modifying the interaction of the detonation wave with the liner. One design uses the placement of an inert-silicone inclusion or multiple inclusions in the HE region to alter the detonation wave shape prior to striking the metal liner of the shaped charge. Another design parameterization that we consider involves making direct modifications to the geometry of the metal liner. 

There are many possible designs that can result from these parameterizations and predicting how these parameterizations affect jet formation is nontrivial. Thus, we use surrogate-based optimization techniques to automatically discover nonintuitive design configurations using 2D hydrodynamics simulations. We fit a neural network surrogate model to hydrodynamics simulation results to quickly assess the performance of various designs. Gradient-based optimization is performed on the surrogate model to find the local optima within the parameter search domain. We also perform experiments using additively manufactured explosive assemblies to produce the unique geometries of the optimal designs to validate our simulations. Overall, the optimal enhancement or suppression designs developed from simulation are effective at altering jet formation and compare well with the corresponding experimental results.

\section{Methodology}\label{sec:methodology}

\subsection{Theory and computational methods}

\subsubsection{Theory of explosively driven RMI jet formation}

\begin{figure}[!htp]
    \centering
    \includegraphics[width=7.0cm]{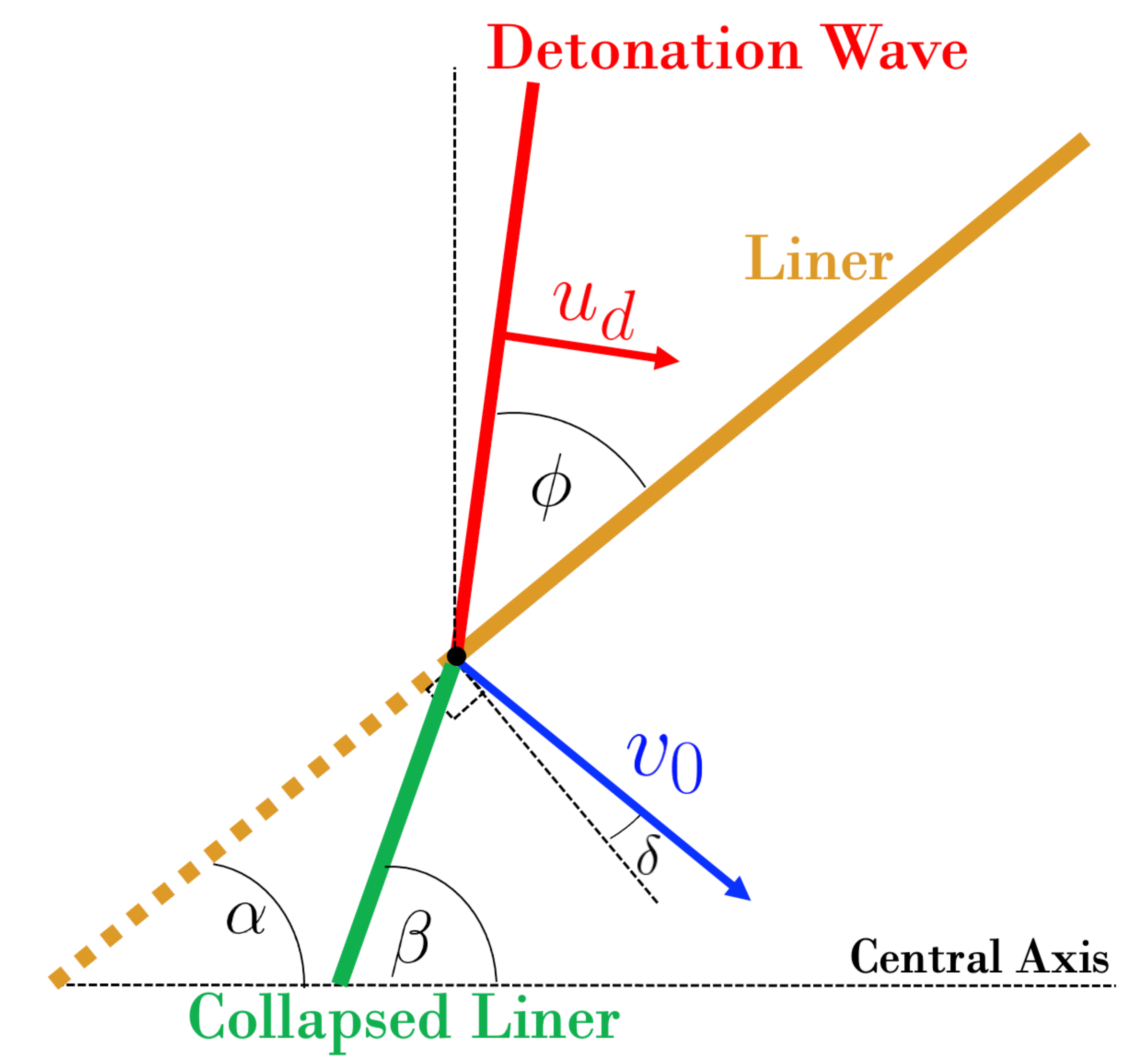}
    \qquad
    \centering
    
    \caption{This diagram shows the collapse of the liner from an initial angle $\alpha$ to the collapse angle $\beta$ as it is struck by a detonation wave moving with velocity $u_{d}$ and at an angle $\phi$ relative to the initial configuration of the liner. Each section of the liner collapses with a velocity $v_{0}$ as the detonation wave sweeps along the liner. Reproduced from Sterbentz et al.~\cite{Sterbentz2023}, with the permission of AIP Publishing.}
\label{fig:Diagram_Birkhoff}
\end{figure}

When a detonation wave strikes the metal cavity liner of a shaped charge, vorticity is deposited along the liner that induces an unstable fluid motion or RMI. The instability forces the liner mass to collapse toward and along the central axis of the shaped charge (see Figure \ref{fig:Diagram_Birkhoff}). This produces a high-velocity jet at the apex of the liner cavity as well as a more slowly moving slug that trails the jet. A general sense for the fluid mechanics involved in the formation of a jet in a shaped charge can be better understood by following the analytical theory developed by Birkhoff et al.~\cite{Birkhoff1948} When a detonation wave moving at velocity $u_{d}$ strikes the liner, vorticity is deposited along the liner and the liner begins to deform. According to the theory of Birkhoff et al., this causes the liner to collapse from an initial angle $\alpha$ to a collapse angle $\beta$ (see diagram in Figure \ref{fig:Diagram_Birkhoff}). For most realistic scenarios, the collapse angle $\beta$ is difficult to determine analytically and is dependent on a number of parameters including $\alpha$, $u_{d}$, the orientation of the detonation wave, as well as material and geometric properties of the liner. The collapse angle $\beta$ will also generally vary along the liner, but is approximated as constant assuming a steady-state case.

As the liner collapses, the liner material is forced into both a high-velocity jet moving with velocity $v_{j}$ that is equal to~\cite{Birkhoff1948,Pugh1952,Sterbentz2023}
\begin{equation} \label{eq:general_v_jet}
v_{j} = v_{0} \csc \bigg(\frac{\beta}{2} \bigg) \cos\bigg( \alpha - \frac{\beta}{2} + \delta \bigg),
\end{equation}
and a lower velocity slug with velocity $v_{s}$ equal to~\cite{Birkhoff1948,Pugh1952}
\begin{equation} \label{eq:general_v_slug}
v_{s} = v_{0} \sec \bigg(\frac{\beta}{2} \bigg) \sin \bigg( \alpha - \frac{\beta}{2} + \delta \bigg).
\end{equation}
A detailed derivation of the above equations is provided in Birkhoff et al.~\cite{Birkhoff1948} and Pugh et al.~\cite{Pugh1952}. In Equations (\ref{eq:general_v_jet}) and (\ref{eq:general_v_slug}), $v_{0}$ is the collapse velocity that describes how quickly a section of the liner collapses from $\alpha$ to $\beta$. Note that, for a realistic jet, $v_{j}$ and $v_{s}$ will generally vary to some degree along the length of the jet or slug, which is ignored in this analysis. 

The collapse velocity $v_{0}$ is not perpendicular to the pre-collapsed liner, but is offset by an angle $\delta$ (see Figure \ref{fig:Diagram_Birkhoff}) that is given by
\begin{equation} \label{eq:Birkhoff_delta}
\delta = \frac{\beta - \alpha}{2}.
\end{equation}
Substituting Equation ($\ref{eq:Birkhoff_delta}$) into Equations (\ref{eq:general_v_jet}) and (\ref{eq:general_v_slug}) produces
\begin{equation} \label{eq:Birkhoff_v_jet}
v_{j} = v_{0} \csc \bigg(\frac{\beta}{2} \bigg) \cos\bigg( \frac{\alpha}{2} \bigg),
\end{equation}
and
\begin{equation} \label{eq:Birkhoff_v_slug}
v_{s} = v_{0} \sec \bigg(\frac{\beta}{2} \bigg) \sin \bigg( \frac{\alpha}{2} \bigg),
\end{equation}
respectively. The collapse velocity $v_{0}$ can be determined using trigonometry and is equal to
\begin{equation} \label{eq:Pugh_collapse_v}
v_{0} = \frac{2 u_{d} \sin\delta} {\sin \phi},
\end{equation}
where $\phi$ is the angle of the detonation wave relative to the liner (see Figure \ref{fig:Diagram_Birkhoff}). From Equations (\ref{eq:Birkhoff_v_jet}) and (\ref{eq:Birkhoff_v_slug}), it is clear that the jet velocity $v_{j}$ and slug velocity $v_{s}$ are directly proportional to $v_{0}$. 

\begin{figure}[!htp]
    \centering
    \includegraphics[width=8.75cm]{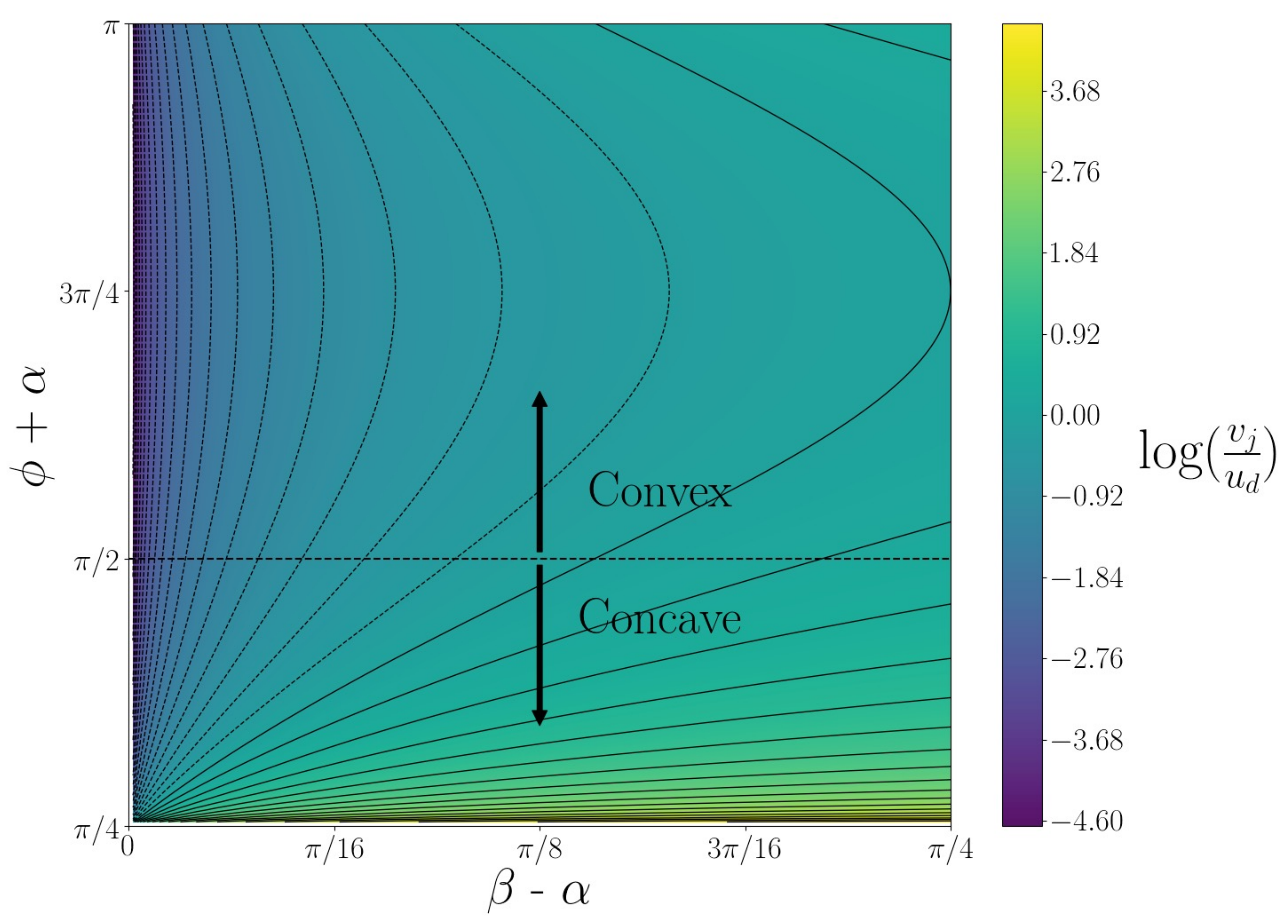}
    \qquad
    \centering
    
    \caption{The jet velocity $\log(v_{j}/u_{d})$ as a function of the detonation wave angle relative to the central axis ($\phi + \alpha$) and the collapse angle relative to the original liner angle ($\beta - \alpha$) in units of radians. The data in this plot is generated using Equation (\ref{eq:Birkhoff_v_jet}) for the case where the original liner angle $\alpha = \pi/4$. This plot shows that concave detonation waves ($0 \leq \phi \leq \frac{\pi}{2} - \alpha$) tend to increase $v_{j}$ relative to a detonation wave moving parallel to the central axis ($\phi = \pi/2 - \alpha$), whereas convex detonation waves ($\frac{\pi}{2} - \alpha \leq \phi \leq \frac{\pi}{2}$) tend to decrease $v_{j}$. }
\label{fig:Birkhoff_plot}
\end{figure}

We now consider three general cases that depend on the detonation wave angle $\phi$ and we assume that the collapse angle $\beta$ remains approximately constant as we vary $\phi$. For the first case, the detonation wave is moving with velocity $u_{d}$ parallel to the central axis (i.e., $\phi = \pi/2 -\alpha$), such that
\begin{equation} \label{eq:case_1}
\phi = \frac{\pi}{2} - \alpha \implies  v_{0} = \frac{2 u_{d} \sin{\delta}}{\cos{\alpha}},
\end{equation}
by substituting $\phi = \pi/2 - \alpha$ into Equation (\ref{eq:Pugh_collapse_v}). For the second case, we consider a ``concave" detonation wave (i.e., $\phi < \pi/2 -\alpha$), which implies
\begin{equation} \label{eq:case_2}
0 \leq \phi \leq \frac{\pi}{2} - \alpha \implies  v_{0} \geq \frac{2 u_{d} \sin{\delta}}{\cos{\alpha}}.
\end{equation} 
Finally, for the third case, a ``convex" detonation wave (i.e., $\phi > \pi/2 - \alpha$) implies that
\begin{equation} \label{eq:case_3}
\frac{\pi}{2} - \alpha \leq \phi \leq \frac{\pi}{2} \implies  v_{0} \leq \frac{2 u_{d} \sin{\delta}}{\cos{\alpha}}.
\end{equation} 
This demonstrates that the collapse velocity $v_{0}$ can be modulated by varying the angle $\phi$ at which the detonation wave strikes the liner to either increase or decrease $v_{j}$ and $v_{s}$. Realistically, the collapse angle $\beta$, and consequently $\delta$, is also a function of $\phi$ and varies as $\phi$ is altered. The jet velocity $v_{j}$ as a function of $\phi + \alpha$ and $\beta - \alpha$ for the case where $\alpha = \pi/4$ (i.e., 45$^\mathrm{o}$ half liner angle) is shown in Figure \ref{fig:Birkhoff_plot}. This figure demonstrates that $v_{j}$ depends on both $\phi$ and $\beta$. Therefore, the relations in Equations (\ref{eq:case_1})--(\ref{eq:case_3}) are only rough approximations used to illustrate the point that the detonation wave angle $\phi$ can have a significant effect on both the jet and slug velocity. 

In addition to $v_{j}$ and $v_{s}$, we must also consider the amount of liner mass that is deposited in the jet and the slug. The portion of the total liner mass $m$ that is advected into the jet $m_{j}$ versus the slug $m_{s}$ is dependent on the collapse angle $\beta$ by the following relations:~\cite{Birkhoff1948,Pugh1952,Sterbentz2023}
\begin{equation} \label{eq:Birkhoff_m_jet}
m_{j} = \frac{m}{2} (1 - \cos\beta),
\end{equation}
\begin{equation} \label{eq:Birkhoff_m_slug}
m_{s} = \frac{m}{2} (1 + \cos\beta).
\end{equation}
As previously mentioned, $\beta$, and consequently $m_{j}$ and $m_{s}$ according to Equations (\ref{eq:Birkhoff_m_jet}) and (\ref{eq:Birkhoff_m_slug}), varies with $\phi$, such that an optimization problem is created where $\phi$ must be carefully selected. The complex interplay between these variables as well as the approximations and simplifications made in the analytical theory have necessitated the use of computational optimization and machine learning methods coupled to hydrodynamic simulations, which we describe in the following sections.

\subsubsection{Setup and parameterizations}
\begin{figure*}[!htp]
    \centering
    \includegraphics[width=18.0cm]{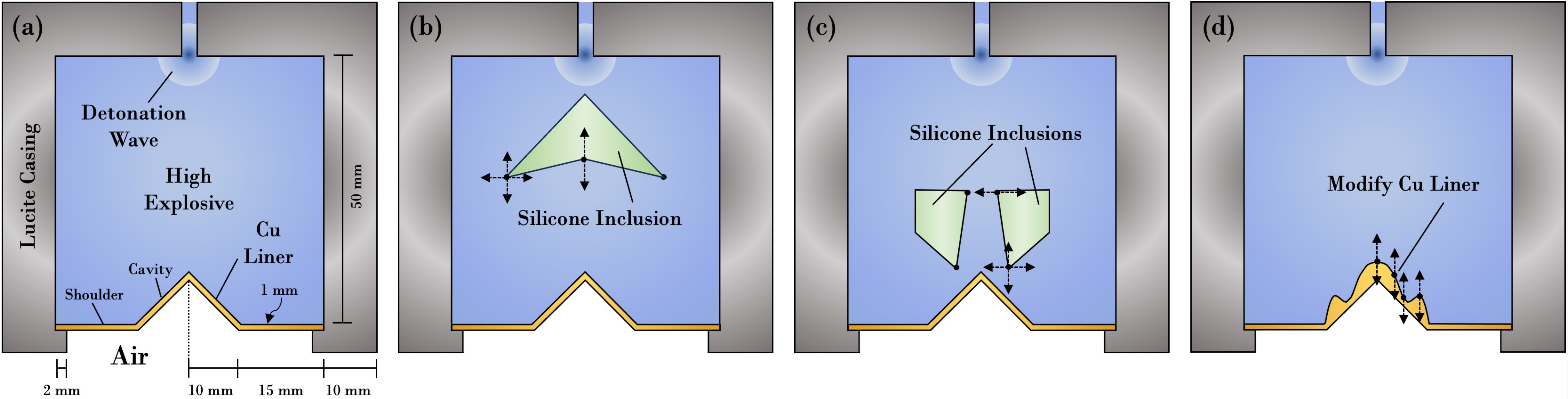}
    \qquad
    \centering
    
    \caption{Baseline design and parameterizations tested in this work. In all cases, the setup consists of HE contained within a Lucite plastic casing. At the front of the HE region is a copper liner with both a cavity region (angled part of liner) and a flat shoulder region. The setup is surrounded by air at ambient conditions. The parameters that are modified in the analysis and in the directional degrees of freedom are represented by the black dots and arrows, respectively. (a) The baseline linear shaped charge setup that we consider for both our simulations and the experiments. (b) A chevron-shaped silicone inclusion in the HE region. (c) Two symmetric silicone inclusions in the HE regions for causing detonation wave diffraction. (d) modification to the contour of the HE/liner interface.}
\label{fig:Diagrams}
\end{figure*}

The linear shaped charge setup that we use in our analysis involves HE (manufactured by Lawrence Livermore National Laboratory) confined in a Lucite plastic casing where the front side of the lucite casing is open. On the front side, a copper liner is used to cover a V-shaped cavity with a 90$^\mathrm{o}$ angle, which seeds the instability that creates the jet. The detonation is initiated at the back side of the casing and produces a detonation wave that traverses the HE region before striking the liner. A cross section of this general setup is shown in Figure \ref{fig:Diagrams}(a) and is used in both our simulations and experiments. To enhance or suppress jet penetration, we have chosen three design parameterizations to focus on for this analysis. Note that the design points that we vary for each parameterization are marked on the diagrams in Figure \ref{fig:Diagrams}(b)--(d) using double-headed arrows to indicate the directions in which the design parameters are altered.

The first design parameterization involves placing a single silicone inclusion within the HE region of the setup as shown in Figure \ref{fig:Diagrams}(b), which we use to enhance jetting. Three design parameters are used that dictate the shape of this silicone inclusion. The silicone is inert in this scenario and has a shock speed that is lower than the detonation velocity of the wave passing through the HE. Thus, the shape of the detonation wave front that strikes the liner can be modified to produce a ``concave" wave front. As previously described, the angle at which the detonation wave strikes the liner can significantly affect jet formation and a concave detonation wave tends to enhance the jet velocity~\cite{Birkhoff1948,Pugh1952,Sterbentz2023}.

The second design parameterization that we consider involves two symmetric silicone inclusions with three design parameters that dictate the shape of these inclusions [see Figure \ref{fig:Diagrams}(c)]. The parameterization in Figure \ref{fig:Diagrams}(c) is used to suppress jetting and has the opposite effect to the parameterization shown in Figure \ref{fig:Diagrams}(b). For the parameterization with two symmetrical inclusions, the detonation waves passes quickly through the narrow HE region between the silicone areas, causing the detonation wave to diffract and creating a ``convex" detonation wave. This convex detonation wave tends to aid in suppressing jetting by reducing the jet velocity as previously discussed~\cite{Sterbentz2023}.

The third design involves modifying the shape of the copper liner [see Figure \ref{fig:Diagrams}(d)] to both enhance and suppress jetting. The design in Figure \ref{fig:Diagrams}(d) uses cubic splines to interpolate between four design points to produce a smooth continuous liner shape in the angled groove region. This parameterization can be used to alter the angle at which the detonation strikes the side of the liner that is contact with the HE and therefore affects the jet velocity. It can also be used to increase or reduce the mass that is advected into the jet versus the slug to enhance or suppress jet penetration.

We use the MARBL hydrodynamics code to construct our simulations, which uses an arbitrary Lagrangian--Eulerian method with high-order finite elements to solve the conservation equations \cite{Anderson2020,Rieben2020,Anderson2018}. MARBL currently has two available HE detonation models: (1) programmed burn and (2) reactive flow. The programmed burn model was developed using Chapman--Jouguet theory and assumes a constant detonation velocity \cite{Handley2018}. While a programmed burn model is useful for a variety of applications, the detonation wave in this model requires that detonated HE be within the line-of-sight of the initial detonation point. This works well for simulating the baseline case in Figure \ref{fig:Diagrams}(a) and the design parameterization in Figure \ref{fig:Diagrams}(d). On the other hand, the reactive flow model developed by Cochran and Tarver \cite{Cochran1979,Lee1980,Cochran1983} does not have this line-of-sight limitation and is needed for simulating the designs in Figure \ref{fig:Diagrams}(b) and (c). However, the Cochran--Tarver reactive flow model that we use has several parameters that are less constrained for the HE and consequently tends to overestimate the jet velocity to some degree. 

\subsubsection{Optimization methodology}\label{sec:opt_method}

The primary objective of our optimization process is to increase or decrease the penetration depth of a linear shaped charge jet into a given material. However, simulating the penetration of a jet into a material can be computationally intensive. For this reason, we have chosen to use a more easily computable metric in our optimization analysis that does not require a full simulation of jet penetration. The following estimate for the penetration depth of a shaped-charge jet into a given material was developed by Birkhoff et al. \cite{Birkhoff1948, Sterbentz2023} for a jet of variable density $\rho_j (x)$ using Bernoulli's equation
\begin{equation} \label{eq:pen_depth}
d = \frac{1}{\sqrt{\rho_{t}}} \int_{0}^{L} \sqrt{\rho_{j}(x)} dx,
\end{equation}
where $\rho_t$ is the density of the material to be penetrated, $L$ is the length of the jet, and $x$ is the coordinate parallel to the length of the jet. To make this relation more general, we can remove the penetrated material density from the equation to produce the following metric
\begin{equation} \label{eq:metric_Bernoulli}
\Phi_{1} = \int_{0}^{L} \sqrt{\rho_{j}(x)} dx.
\end{equation}

As described in Sterbentz et al.~\cite{Sterbentz2023}, the above metric is generally not accurate enough to predict penetration depth in our simulations and does not account for density variations in the $y$-direction (i.e., perpendicular to the length of the jet). We opt for the alternative metric that is related to the kinetic energy of the jet \cite{Sterbentz2023}
\begin{equation} \label{eq:metric_KE}
\Phi_{2} = \frac{1}{2} \int_{-a}^{a} \int_{0}^{L}\rho_{j}(x,y) v_{j}(x,y)^{2} dx dy,
\end{equation}
where $v_{j}$ is the velocity of the jet and $a$ is some distance in the $y$-direction from the centerline of the jet (we use $a\approx0.1$~cm in our analysis). The second integral accounts for variations in density and velocity from the centerline of the jet. This metric correlates well with jet penetration \cite{Sterbentz2023} and we subsequently use this metric in our analysis. The metric in Equation (\ref{eq:metric_KE}) ensures that both the velocity and mass of the jet is sufficiently high, which are both important for penetration.

To optimize our objective metric in Equation (\ref{eq:metric_KE}), we use the optimization methodology outlined in Sterbentz et al.~\cite{Sterbentz2023} This involves first running several hundred hydrodynamic simulations using a Latin hypercube sampling method to choose a set of design parameter vectors $\bold{z}$ within the bounds of our parameter search domain $\mathcal{Z}$. Note that these simulations are 2D to preserve disk space and reduce computational time, although we use full 3D simulations for our final comparison with the experiments.

A surrogate model of the design space is then generated by training a deep neural network (DNN) with Gaussian activation functions, two hidden layers, and between 50 to 100 nodes per layer ~\cite{Queipo2005,White2019,Pytorch}. This surrogate model acts as a function that we denote as $\hat{\Phi}(\textbf{z})$, which describes the relation between the input design parameters $\textbf{z}$ and the objective metric that we are trying to optimize [i.e., $\Phi_{2}$ in Equation (\ref{eq:metric_KE})]. The limited-memory Broyden--Fletcher--Goldfarb--Shanno (LBFGS) gradient-based method~\cite{} is repeatedly performed on the DNN surrogate model with random initial points to solve the following optimization problem for the suppression case:
\begin{equation} \label{eq:min_suppress}
\displaystyle \min_{\textbf{z} \in \mathcal{Z}} \hat{\Phi}(\textbf{z}),
\end{equation}
and the following optimization problem for the enhancement case:
\begin{equation} \label{eq:min_suppress}
\displaystyle \min_{\textbf{z} \in \mathcal{Z}} -\hat{\Phi}(\textbf{z}).
\end{equation}
Ideally, this repeated use of the LBFGS algorithm should locate all local minima (or maxima) in the surrogate model design space. Additional simulations are then performed near the best local minima to narrow in on an optimal point.

\subsection{High explosives detonation experiment}

\subsubsection{Experiment design}
To better compare the proposed design parameterizations and their effect on RMI jet formation to previous work, the experimental platform for the detonation experiments is adapted from that used in Kline and Hennessey et al.~\cite{Kline2024} This involves using a linear shaped charge with the general layouts shown in Figure \ref{fig:Diagrams} and flash X-ray (FXR) radiography to produce images during and after detonation. It is important to note that the experiment design may yield images with parallax effects for any off-center images (as is the case in late-time images of these experiments). Photonic doppler velocimetry (PDV) is also used to recover velocity traces for both jetting in the cavity region and the shoulder region. The total mass of high explosives used in the experiment is limited to 250 g to minimize the likelihood of damage to the testing facilities and multiple 2.54 cm thick steel plates are evenly spaced a distance of 2.54 cm apart to prevent penetration into the surrounding tank enclosure. The explosive charges have outer dimensions of 50 mm x 50 mm x 50 mm and have a 20 mm-wide triangular prism cutout where the jet forms. At the top of the explosive charge is a single line wave generator (LWG) spanning the entire depth of the charge and running parallel to the cutout. The explosive charge is housed in a 10 mm thick casing that is necessary for part fabrication. The baseline design and designs that modulated jet velocity with silicone inclusions feature a 1 mm thick copper liner with a 90$^\mathrm{o}$ liner angle. The designs that modulate the jet velocity by modifying the copper liner/explosive interface shape have 1 mm thick shoulders, but have variable thickness across the test section. A detailed drawing of the 3D baseline experiment can be seen in Kline and Hennessey et al.~\cite{Kline2024}

\subsubsection{Materials and part fabrication}
Considering that the designs generated via the machine learning methods can be complex, a direct-ink-write additive manufacturing approach is used to fabricate the parts. In all experiments, an extrudable high-power explosive (Lawrence Livermore National Laboratory) is used to drive the metal liners. Two experiments feature inert silicone inclusions made of DOWSIL SE1700 (Dow, Inc.) colored with blue SilcPig pigments (Smooth-On) to visually distinguish materials during the printing process. Prior to printing, both materials are mixed and degassed with a planetary centrifugal mixer (Flacktek DAC 600.2 SpeedMixer). Materials are then transferred to cartridges (Nordson EFD) and placed back in the mixer for a second time. After the material is spun down, a piston is added to the cartridges and they are placed into pressurized retainer systems mounted to the 3D printer. Oxygen-free, high conductivity copper liners are procured and used as-received from the manufacturer (Protolabs). The detonation event is initiated using an RP-2 exploding bridge wire detonator (Teledyne Defense Electronics) which initiates the hand-packed LWG filled with a PETN-based explosive (Lawrence Livermore National Laboratory). 

Five components are fabricated for this experiment series: one monolithic HE charge on a standard liner (baseline), two components with silicone inclusions in the HE on a standard liner, and two components with monolithic HE charges on modified liners. The extrudable explosive is space-filling and therefore spreads after being dispensed, thus necessitating printing into a mold (3D printed, Stratasys), which doubles as a casing. Materials are printed directly into the molds and onto the liners to ensure good contact between the HE and metal. The additive manufacturing process has been previously described in Kline and Hennessey et. al~\cite{Kline2024} and is only briefly discussed here. The components are prepared via a direct-ink-writing process using a custom-built 3D printer that runs internally-developed software to synchronize motion and material flow.~\cite{Kline2024,Grapes2021} The printer is comprised of an Aerotech 3D motion system (Aerotech PRO165LM, Aerotech PRO165SL) driven by an Aerotech Npaq running A3200 software motion controller. Materials are dispensed separately from a pressure-fed material reservoir into progressive cavity pumps (Viscotec preeflow eco-PEN450) and extruded through a straight-barrel nozzle (Nordson EFD). High-density infill toolpaths are prepared for the components using commercial slicing software (Simplify3D). The toolpath for the HE has a layer height and filament width of 1.0 mm and 1.85 mm, respectively and the toolpath for the silicone (co-printed) has a layer height and filament width of 0.5 mm and 1.27 mm, respectively. After being printed, parts are thermally cured in an oven at 50 $^\mathrm{o}$C overnight. The initiation train (detonator and LWG) are added to the assemblies after the top surface is trimmed flat. 

\subsubsection{Detonation experiment and diagnostics}
As previously discussed, the primary diagnostics for the detonation experiment are FXR radiography and PDV. FXR radiography produces images that can be used to evaluate detonation front and material deformation. However, a limited number of images can be taken during the experiment due to infrastructure limitations. PDV is used as a complementary diagnostic to recover velocity traces at specific points for longer durations. Note that although PDV traces are particularly useful for extracting velocity data, capturing information on a small jetting region (like those formed during shaped charge detonations) is particularly difficult and can have limited utility.  

The detonation experiment was conducted in the Lawrence Livermore National Laboratory's High Explosives Applications Facility spherical tank. The FXR system is comprised of three 450 kV pulsers (L3 Harris) positioned on the exterior with 8$^\mathrm{o}$ radial spacing. The X-rays are collimated with a collimator on the tank walls and on the steel shrapnel catcher, then pass through the experiment region, and are finally captured on a protected digital imaging plate. As the experiment is a linear shaped charge and the FXR heads are placed radially, there is a residual parallax effect in the images as will be seen in the radiographs that we present in Section \ref{sec:results_discussion}. Acquisition time of the FXR images is preprogrammed to capture events of interest, notably early deformation of the detonation front and metal liners and late time images to estimate velocity. The first image to be acquired is perfectly aligned with the experiment (0$^\mathrm{o}$) with the second and third images being captured at -8$^\mathrm{o}$ and +8$^\mathrm{o}$, respectively. Jet velocities based on the FXR images are estimated using ImageJ.~\cite{ImageJ} 

An array of 11 collimating PDV probes (OzOptics LPC-01-1550-9/125-S-0.95-5AS-60-3A-3-5) are placed ~60 mm from the flat shoulder of the copper liner output face and acquired data is used to estimate velocimetry data. Five probes are placed along the centerline of the output face of the shaped charge parallel to the jet forming region and 3 probes are placed on either side of this facing the flat face. A 1550 nm laser (IPG photonics) is used as the light source for the PDV technique and data is recorded using a GHz oscilloscope (Tektronix DPO73304DX). Velocimetry data is extracted using LLNL's EDGE software.~\cite{EDGE}

\section{Results and Discussion}\label{sec:results_discussion}

\subsection{Hydrodynamic simulations}

Using the optimization methodology described in Section \ref{sec:opt_method}, we develop four designs that either enhance or suppress jetting relative to the baseline case that we consider. For each of these designs, we conduct an experiment and a 3D simulation for comparison. We use 3D simulations to account for any 3D phenomena not captured by the 2D simulations and for a better comparison with the FXR radiography images that may include parallax effects. In this section, we provide 3D simulation results and compare these to the FXR radiography images from the corresponding experiments. While jet velocity is important for evaluating the effectiveness of a shaped-charge jet, the concentration of the mass is another important jet characteristic, as it affects the kinetic energy and penetrative properties of the jet. However, the experiment diagnostics tend to measure velocities, so we focus primarily on comparing velocities in the following discussion of our results.

\begin{figure*}[!htp]
    \centering
    \includegraphics[width=14cm]{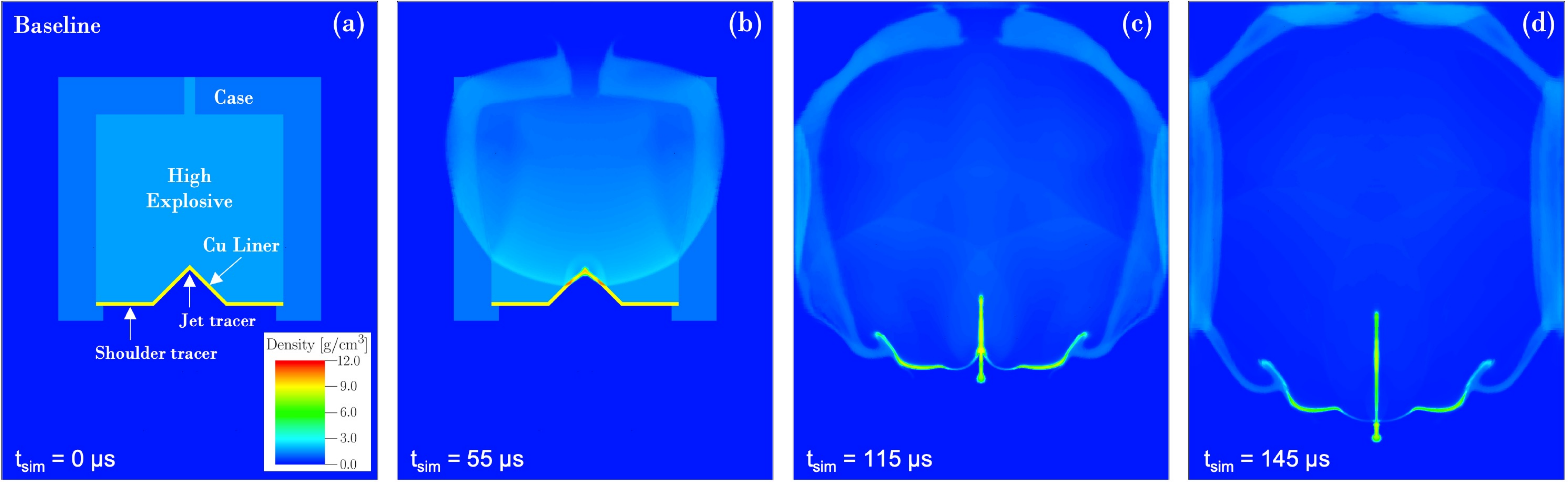}
    \qquad
    \centering
    
    \caption{3D simulation images at several time instants for the baseline case, where the initial detonation occurs at time $t = 0 \; \mathrm{\mu s}$: (a) $t < 0 \; \mathrm{\mu s}$; (b) $t = 6 \; \mathrm{\mu s}$; (c) $t = 12 \; \mathrm{\mu s}$; (c) $t = 15 \; \mathrm{\mu s}$.}
\label{fig:sim_baseline}
\end{figure*}

Before investigating any methods to modulate jet velocity, we ran simulations on the ``baseline" case. As previously described, this component simply consists of the monolithic HE printed directly onto the copper liner placed within the casing and initiated with a line wave generator. Figure \ref{fig:sim_baseline} shows density plots from simulation at several time instants for the baseline case. The jet and shoulder velocities of the baseline case were estimated to be 5.23~km/s and 3.59~km/s, respectively, and we use these values to compare the effect that our designs have on the jet velocities.

\begin{figure*}[hbt!]
    \centering
    \includegraphics[width=14cm]{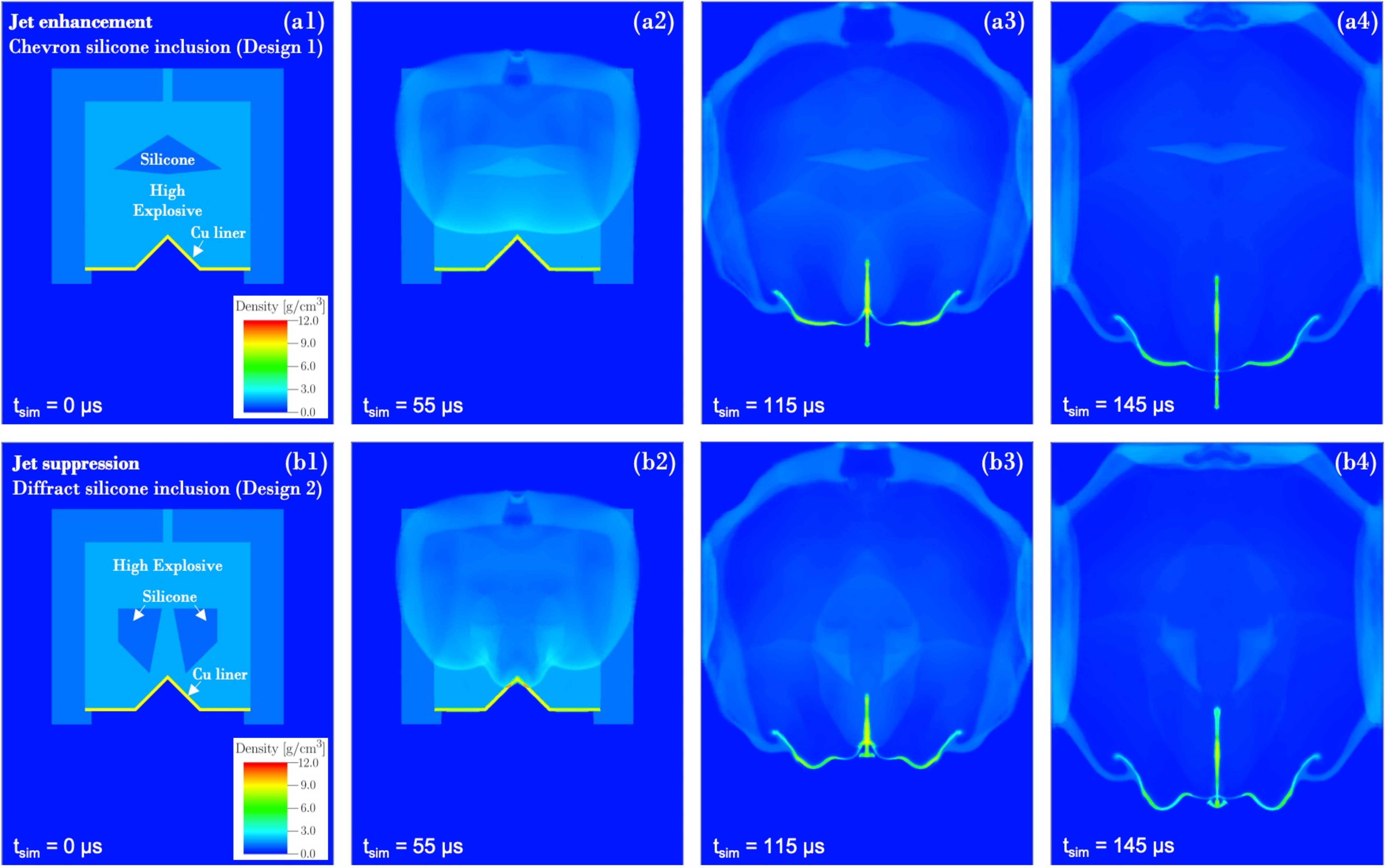}
    \qquad
    \centering
    
    \caption{3D simulation images at several time instants for designs with silicone inclusions: (a1--a4) Design 1, chevron silicone inclusion design for enhanced jetting and (b1--b4) Design 2, diffraction silicone inclusion for suppressed jetting. Initial detonation occurs at time $t = 0 \; \mathrm{\mu s}$.}
\label{fig:sim_silicone}
\end{figure*}

The first parameter space we explore in this work uses silicone inclusions to modulate the RMI jet velocity. Unlike previous work performed in Kline and Hennessey et al.~\cite{Kline2024}, these silicone inclusions are ``suspended" in the HE to shape the detonation wave front before reaching the liner material. Figure \ref{fig:sim_silicone}(a) shows the results for the jet enhancement design, which features a chevron-shaped silicone inclusion (Design 1). The resulting jet velocity and shoulder velocity in the 3D simulations of the components is 6.18~km/s and 3.73~km/s, respectively. Design 1 has the highest jet velocity of all the designs presented in our analysis. For this design, the silicone inclusion slows the velocity of the detonation wave in the central region before striking the liner. This gives the detonation wave a flatter and slightly ``concave" shape that affects the angle at which the wave strikes the liner [see Figure \ref{fig:sim_silicone}(a2)]. A different modification involving two silicone inclusion components can also be used to reduce the jet velocity. A diffraction silicone inclusion design (Design 2) is presented in Figure \ref{fig:sim_silicone}(b) which has peak jet and shoulder velocities of 4.65~km/s and 3.62~km/s, respectively. Design 2 provides the opposite effect to Design 1. The two silicone inclusions reduce the detonation wave velocity away from the central axis. Design 2 also creates a diffraction effect as the detonation wave passes between the silicone inclusions, which produces a highly curved or ``convex" detonation wave in the region of the liner cavity [see Figure \ref{fig:sim_silicone}(b2)]. This diffracted detonation wave moves nearly perpendicular to the back surface of the liner cavity and seems to be a reasonably effective method for suppressing the jet.

\begin{figure*}[hbt!]
    \centering
    \includegraphics[width=14cm]{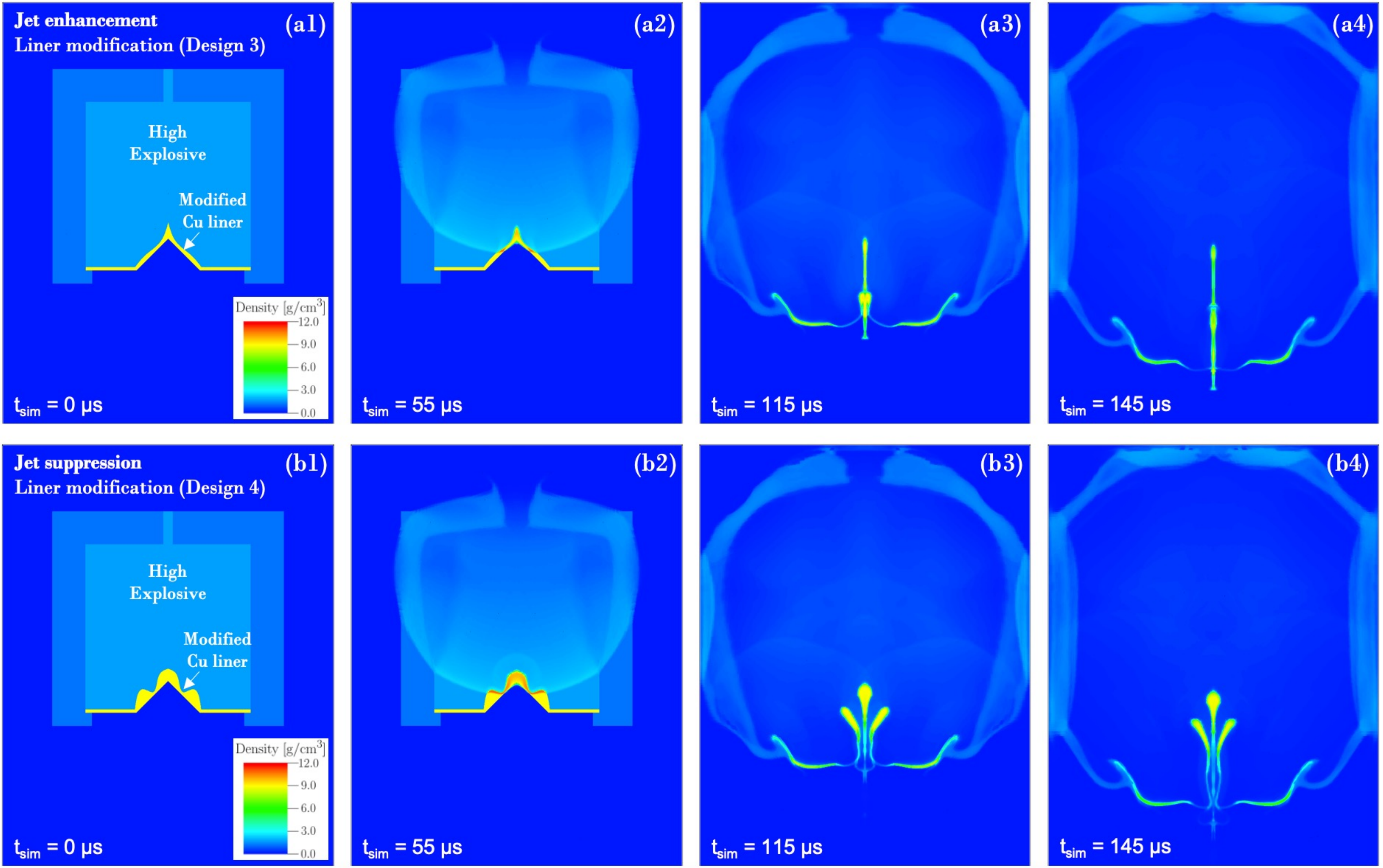}
    \qquad
    \centering
    
    \caption{3D simulation images at several time instants for designs with liner geometry modifications: (a1--a4) Design 3, liner modification for enhanced jetting and (b1--b4) Design 4, liner modification for suppressed jetting. Initial detonation occurs at time $t = 0 \; \mathrm{\mu s}$.}
\label{fig:sim_liner}
\end{figure*}

The second set of variations we made on the explosive components sought to modify the jet velocity by varying the contour of the HE/liner interface in the angled cavity region. A set of cubic splines is varied in two separate optimization routines to yield different designs for enhancing and suppressing jetting. Design 3, the modified liner design which enhances jet velocity, can be seen in Figure \ref{fig:sim_liner}(a). This design features a slightly thicker liner near the cavity region and a prominent peak right behind the apex of the liner cavity. Design 3 yields jet velocities of 5.02~km/s and shoulder velocities of 3.62~km/s. Despite having lower jet velocity than Design 1, Design 3 has more mass concentrated along the central axis due to the modified shape of the copper liner shown in Figure \ref{fig:sim_liner}(a), which indicates that the kinetic energy is likely high as well. The additional liner mass concentrated around the apex of the liner appears to increase the amount of mass that is deposited in the high-velocity jet without significantly modifying the interaction between the detonation wave and the liner. Design 4, the suppression case involving a modified liner, seems to have the lowest jet velocity, which is comparable to the shoulder velocity [see Figure \ref{fig:sim_liner}(b)]. In fact, there does not appear to be a coherent jet for Design 4, although some very low-density, high-velocity ``dust" is observed in the simulation. Design 4 instigates the advection of the liner mass into large slugs due to the wavy profile of the liner, indicating that much of the additional liner mass is being significantly slowed.

\subsection{Part fabrication and detonation experiments}

Having determined optimal designs for both enhancing and suppressing RMI jetting in the linear shaped charge analogues, components are fabricated using an additive manufacturing process. Part fabrication requires only minor variations to flow rates of materials to maintain part quality over the duration of the print. After printing and thermal curing, components are inspected using an X-ray computed tomography (XRCT) to evaluate part quality with the same instruments described previously in Kline and Hennessey et al.~\cite{Kline2024} The XRCT analysis revealed that, although the HE material itself only had minor voids introduced during the curing process, there are some larger voids that were found at the HE/silicone interfaces and perimeter/infill interfaces for the multi-material components. This likely stems from trapped air in the silicone (which remains after a multi-step degassing process) migrating at high temperatures and merging, thus creating minor separations between different features. After evaluating these defects, it was determined that they were minor enough to proceed with detonation testing without dramatically impacting the quality of the results.

\begin{figure*}[hbt!]
    \centering
    \includegraphics[width=16cm]{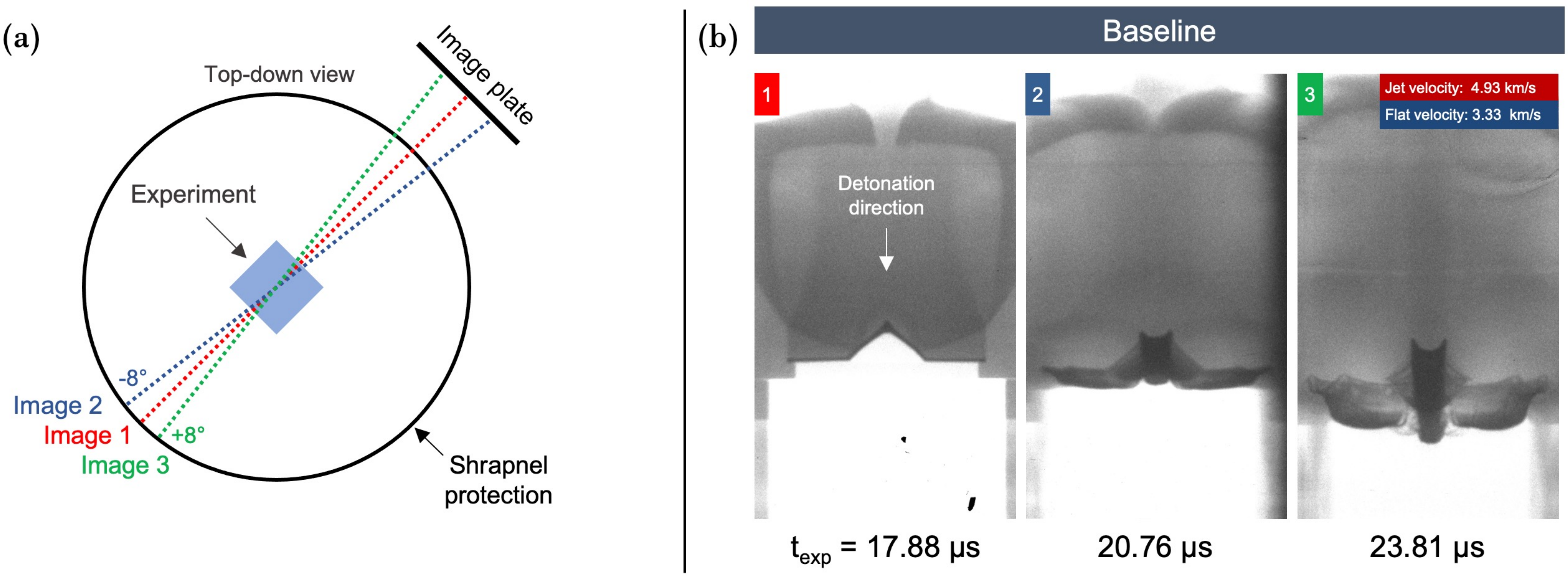}
    \qquad
    \centering
    
    \caption{(a) Experiment layout and (b) experimental radiographs for the baseline design. The orientation of the image source/plate to the experiment is 0$^\mathrm{o}$, -8$^\mathrm{o}$, and +8$^\mathrm{o}$ from left to right. Note that the spots in image 1 are from a damaged imaging plate.}
\label{fig:exp_layout}
\end{figure*}

The primary diagnostic that we use to evaluate component performance in these experiments is FXR radiography. FXR radiography is particularly well suited for these experiments as it can be easily used to distinguish materials based on their intrinsic properties (e.g. density) and can capture a full-field image of the experiment. However, as previously described, the existing setup for the FXR capability in LLNL's High Explosives Application Facility does not enable coaxial imaging from the same port and three separate ports with different heads must be used to image the component during the experiment. Figure \ref{fig:exp_layout}(a) depicts the layout of the experimental facility with the shrapnel capture as viewed from the top; three images are to be captured with the first image being taken perfectly aligned to the part to view early-time wave front and material deformations and later images being taken from $\pm$8$^\mathrm{o}$, although this would yield a distortion in the image due to the parallax effect. Static images of each experiment prior to detonation are presented in Appendix \ref{sec:PDV}. We use the VisIt software~\cite{VisIt2012} to create simulated FXR radiograph images using output data from our simulations for comparison to the FXR images captured in the experiment. 

PDV data was also captured for each experiment performed, but this was considered a secondary diagnostic to the FXR images. Capturing velocities of jets is particularly challenging for PDV as minor misalignments of the fiber probes or jet deviations from perfect trajectories can result in poor signal-to-noise ratios or complete loss of signal. PDV data for all experiments is presented in Appendix \ref{sec:PDV} for completeness.

The first experiment that we performed is the baseline design consisting of a monolithic charge of HE against the standard liner (HE mass $\approx$209 g). FXR radiographs from the first experiment can be seen in Figure \ref{fig:exp_layout}(b). Velocities of the jetting and shoulder regions of the liner are estimated using ImageJ.~\cite{ImageJ} Similar to the identical experiment performed in Kline and Hennessey et al.~\cite{Kline2024}, this experiment clearly demonstrates the desired jetting behavior and had similar terminal jet and shoulder velocities of 4.93~km/s and 3.33~km/s, respectively.~\cite{Kline2024} A more detailed discussion on the results expected from the baseline experiment is described in Kline and Hennessey et al.~\cite{Kline2024} These results slightly deviated from the hydrodynamic simulations which predicted a late-time velocity of 5.23~km/s and 3.59~km/s for the jet and shoulder regions, respectively. These differences are relatively minor and may be attributable to deviations in the material models used in the simulations from the actual materials as well as small defects in the 3D-printed materials used in the experiments. PDV data collected from the experiment also closely match the predicted results from hydrodynamic simulations (Figure \ref{fig:PDV_baseline}).

\begin{figure*}[hbt!]
    \centering
    \includegraphics[width=16cm]{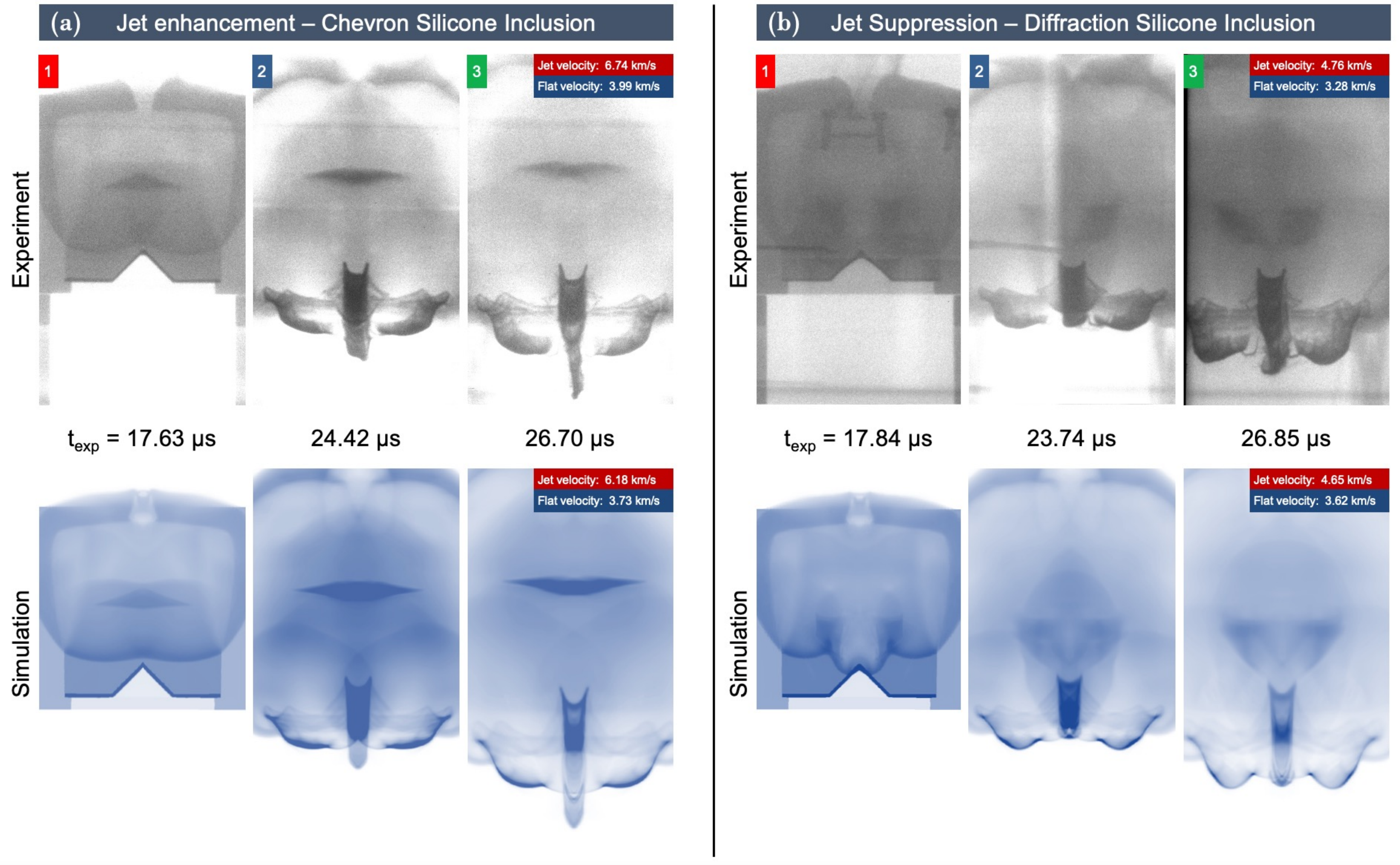}
    \qquad
    \centering
    
    \caption{A comparison between experimental and simulated FXR radiographs for designs with silicone inclusions. In both figures, the top row is the experimental radiograph and the lower row is a simulated radiograph. In all sets, the orientation of the image source/plate to the experiment is 0$^\mathrm{o}$, -8$^\mathrm{o}$, and +8$^\mathrm{o}$ from left to right. (a) Design 1 featuring a chevron silicone inclusion for jet enhancement. (b) Design 2 featuring diffraction silicone inclusions for jet suppression.}
\label{fig:exp_silicone}
\end{figure*}

The second set of experiments focuses on designs which use silicone inclusions as a mechanism to shape the detonation wave fronts to either enhance or suppress jetting during the detonation event. Both sets of radiographs for the jet enhancement (Design 1) and suppression (Design 2) experiments with silicone inclusions can be seen in Figure \ref{fig:exp_silicone}(a) and (b), respectively, where experimentally obtained radiographs are presented alongside the predicted radiographs from simulation below them. The first image for Design 1 [Figure \ref{fig:exp_silicone}(a)] clearly demonstrates a concave detonation wave front induced by the placement of the silicone inclusion. This experimental radiograph almost perfectly matches with the predicted radiograph from the hydrodynamic simulations, which also features this deformation of the wave front. Later images of the experiment highlight the distance the liner travels between two time steps and image analysis revealed that the jet and shoulder regions reached a velocity of 6.74~km/s and 3.99~km/s, respectively. This represents a $\approx$37$\%$ increase in the jet velocity and 19.9$\%$ increase in the shoulder velocity despite a $\approx$8$\%$ reduction in energy content for Design 1. It is important to note that the third experiment image of Figure \ref{fig:exp_silicone}(a) depicts the jet beginning to drift right, likely due to a defect in the component or assembly; velocity for the jet at this late time is only measured based on distance traversed in the vertical direction of the image. Unfortunately, PDV data for the jetting region of the experiment had a poor signal to noise ratio and did not yield usable results. However, the PDV shoulder velocity of the experiments is in relative agreement with those estimated from the images [see Figure \ref{fig:PDV_chevron}(b)].

The diffraction silicone inclusion design (Design 2), which intended to suppress RMI jetting, is able to reduce jet velocity to some degree. The radiographs for the Design 2 experiment at early times demonstrated a clear shock front detonation which corresponded closely with the simulated radiographs for the experiment. Later time comparisons between the experiment and simulation are in relative agreement, however the jet began to deform at late times again. Captured radiographs from this experiment also have some uneven brightness in the images introduced by slight misalignment of a collimator to the FXR head during resetting of the testing facility. The jet and shoulder velocities estimated by the image were 4.76~km/s and 3.28~km/s respectively. This represents a 3.4$\%$ decrease and 1.3$\%$ decrease in jet velocity and shoulder velocity, respectively. However, the energetic content of the experiment decreased by $\approx$14.80$\%$. These results from the FXR analysis are also corroborated by those captured by PDV (Figure \ref{fig:PDV_diffract}), which had adequate signals for measurement over the entire jetting region. In future efforts, this design may need to be altered by replacing or reshaping the inclusions to further aid in suppressing jetting.

\begin{figure*}[hbt!]
    \centering
    \includegraphics[width=16cm]{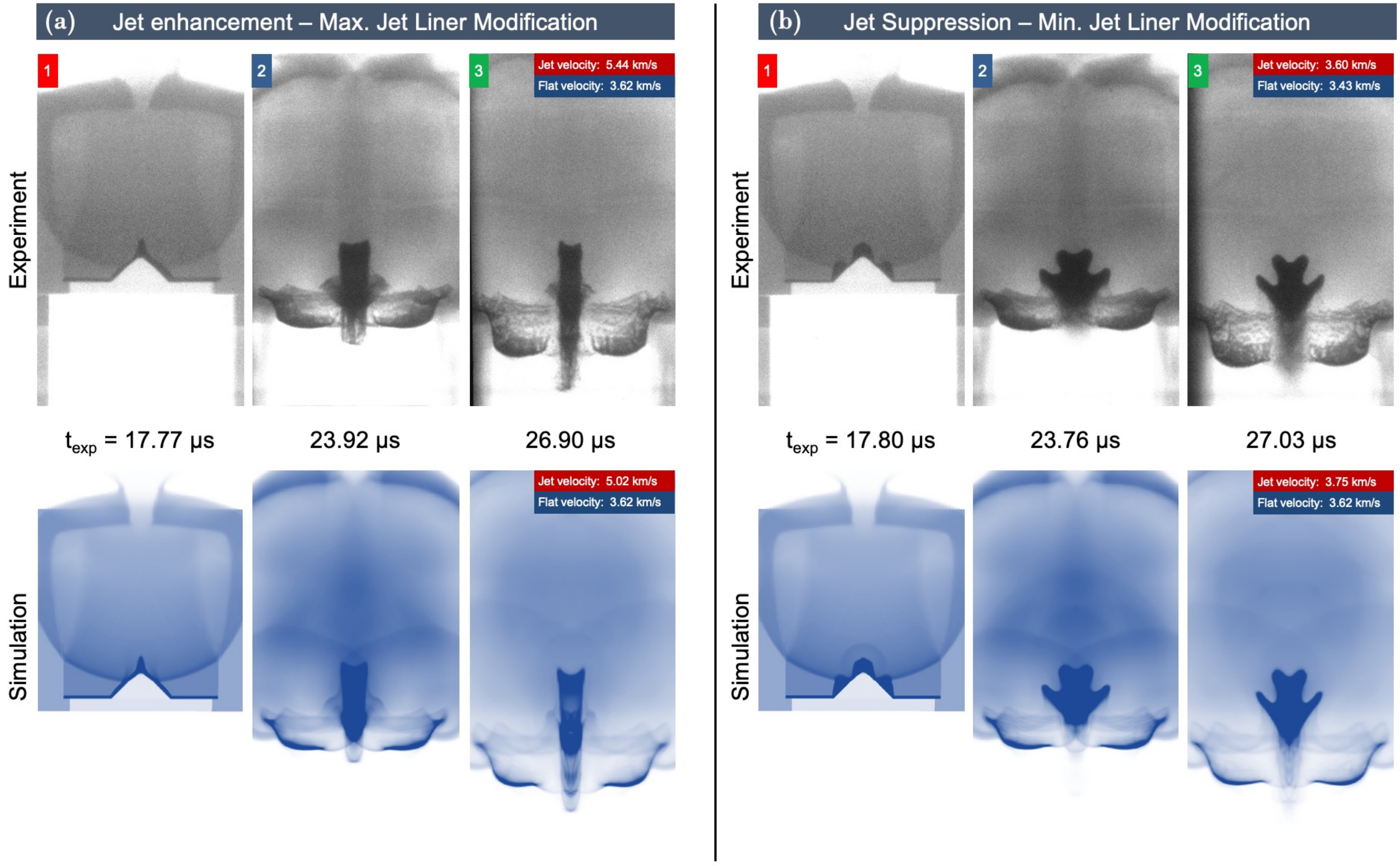}
    \qquad
    \centering
    
    \caption{A comparison between experimental and simulated FXR radiographs for designs with liner geometry modifications. In both figures, the top row is the experimental radiograph and the lower row is a simulated radiograph. In all sets, the orientation of the image source/plate to the experiment is 0$^\mathrm{o}$, -8$^\mathrm{o}$, and +8$^\mathrm{o}$ from left to right. (a) Design 3 featuring a liner modification for jet enhancement. (b) Design 4 featuring a liner modification for jet suppression.}
\label{fig:exp_liner}
\end{figure*}

The last set of experiments conducted in this series focused on modifying the copper liner to enhance or suppress the jet. Figure \ref{fig:exp_liner}(a) and (b) shows the radiographs captured during the jet enhancement (Design 3) and suppression (Design 4) experiments, respectively. Since there are no inert inclusions in the HE, the deformation of the detonation front is not visible prior to the interface between the liner and the explosives, thus the early time images closely resemble those of the baseline case. However, the late time images from these experiment demonstrate the dramatic impact that modifications of the liner may have on the jetting behavior of these components. The estimated jet velocity of Design 3, which intended to increase jet velocity, is 5.44~km/s and the shoulder velocity is 3.62~km/s based on the FXR images. These velocities are 10.4$\%$ and 8.8$\%$ higher than those recorded in the baseline case. Meanwhile, the energetic content of the component decreased by $<1\%$ since the modification to the liner itself was relatively minor. Images captured for the jet enhancement case (Design 3) also closely match the simulated radiographs, as can be seen in Figure \ref{fig:exp_liner}(a). Similar to earlier experiments, the PDV data recorded for this experiment lost signal relatively early before the detonation event was over (see Figure \ref{fig:PDV_liner_max}).

The last design tested (Design 4) has the most profound effect on jet suppression of any of the designs in this set of experiments. The liner modification that intended to suppress jetting appears to entirely suppress the jetting behavior and several large slugs are formed in the wake of the detonation (Figure 7b). Late time images of the detonation event shows low density ``dust" where a jet would have otherwise formed, but in comparison to other experiments, this carries relatively little momentum. The estimated jet and shoulder velocities from the FXR images is 3.60~km/s and 3.43~km/s, respectively. This is a 27.0$\%$ decrease in the jet velocity compared to the baseline case, but only a 2.9$\%$ decrease in the shoulder velocity and 1.9$\%$ reduction in energetic content. Simulated radiographs from this experiment are also in very close agreement with the experimental images. PDV data recorded for this data set was in close agreement with simulations at early times, but the low density dust formed during the detonation event likely scattered any measurable light in the jetting region and resulted in a loss of signal at later times (see Figure \ref{fig:PDV_liner_min}). Design 4 is the most promising method to reduce jetting behavior in this configuration and is one of the simpler designs to manufacture. 

The overall results of both the jet enhancement and suppression experiments with silicone inclusions and liner modifications demonstrate that shock front deformation and detonation wave interactions with the liner can be used to modulate jet velocities using designs driven by machine learning. A summary of the velocities at the jet tip and along the liner shoulder region from simulation tracer data and estimated experiment data are shown in Table \ref{tab:exp_velocities}. The experiment jet and shoulder velocities are estimated directly from the FXR radiography images. Table \ref{tab:exp_velocities} also shows the percent difference between the simulation and the experiment for both the jet and shoulder velocity. In general, the percent difference is less than 10$\%$, which demonstrates good correlation between simulation and experiment. Table \ref{tab:exp_percent} summarizes the percent decrease in the HE mass (i.e., the energetic content) as well as the percent increase or decrease in the jet and shoulder velocities from the baseline design for both simulation and experiment. We also note that the percent difference in the jet velocity between the most enhanced and the most suppressed design is a difference of 87\%, indicating that these designs provide a significant amount of modulability.

\setlength{\extrarowheight}{7pt}
\begin{table*}
\caption{\label{tab:exp_velocities} Comparison between late time jet and shoulder velocity from simulation tracers and estimated from experiment FXR radiography images.}
\begin{tabular}{ >{\centering\arraybackslash}p{1.75cm}|>{\centering\arraybackslash}p{1.7cm}|>{\centering\arraybackslash}p{1.7cm}|>{\centering\arraybackslash}p{1.7cm}|>{\centering\arraybackslash}p{2.2cm}|>{\centering\arraybackslash}p{2.2cm}|>{\centering\arraybackslash}p{1.9cm} }
 Design & Jet (sim.)  [km/s] & Jet (exp.) [km/s] & Jet, $|\% \; \mathrm{Diff.} |$ & Shoulder (sim.)  [km/s] & Shoulder (exp.) [km/s] & Shoulder, $|\% \; \mathrm{Diff.} |$ \\
 \hline
 \hline
 Baseline & 5.23 & 4.93 & 6.18\% & 3.59 & 3.33 & 8.00\%  \\[0.0cm]  
 1 (enhance)  & 6.18 & 6.74 & 8.35\% & 3.73 & 3.99 & 6.62\% \\[0.0cm]  
 2 (suppress) & 4.65 & 4.76 & 2.34\% & 3.62 & 3.28 & 10.2\% \\[0.0cm]  
 3 (enhance) & 5.02 & 5.44 & 7.79\% & 3.62 & 3.62 & 0.09\% \\[0.0cm] 
 4 (suppress) & 3.75 & 3.60 & 4.04\% & 3.62 & 3.43 & 5.65\% \\[0.0cm] 
\end{tabular}
\end{table*}

\setlength{\extrarowheight}{7pt}
\begin{table*}
\caption{\label{tab:exp_percent} The percent increase or decrease (negative) in jet and shoulder velocity from the baseline design for both simulation and experiment.}
\begin{tabular}{ >{\centering\arraybackslash}p{1.75cm}|>{\centering\arraybackslash}p{1.65cm}|>{\centering\arraybackslash}p{1.7cm}|>{\centering\arraybackslash}p{1.7cm}|>{\centering\arraybackslash}p{1.7cm}|>{\centering\arraybackslash}p{2.2cm}|>{\centering\arraybackslash}p{2.2cm}}
 Design & HE mass [g]& HE mass [\%~inc.] & Jet (sim.) [\%~inc.] & Jet (exp.) [\%~inc.] & Shoulder (sim.) [\%~inc.] & Shoulder (exp.) [\%~inc.]  \\
 \hline
 \hline
 Baseline & 208.8 & 0.0\% & 0.0\% & 0.0\% & 0.0\% & 0.0\% \\[0.0cm]  
 1 (enhance) & 191.7 & -8.2\% & 25.4\% & 36.8\% & 3.7\% & 19.9\%  \\[0.0cm]  
 2 (suppress)& 177.9 & -14.8\% & -5.6\% & -3.4\% & 0.7\% & -1.3\%  \\[0.0cm]  
 3 (enhance) & 207.9 & -0.4\% & 1.8\% & 10.4\% & 0.8\%  & 8.8\%  \\[0.0cm] 
 4 (suppress) & 204.8 & -1.9\% & -24.0\% & -26.9\% & 0.7\%  & 2.9\%  \\[0.0cm] 
\end{tabular}
\end{table*}

\section{Conclusions}
We have investigated several design parameterizations for both enhancing and suppressing RMI jetting in linear shaped charges. We used simulations to select optimal designs from these parameterizations and have performed physical experiments of these designs. The optimal designs developed using computational optimization methods are sometimes nonintuitive. However, we have demonstrated through simulation and experiment that these optimal designs are quite effective at controlling jetting. The experiments matched well with the results predicted through simulation, as shown through comparisons between experiment FXR radiography images and simulated X-ray images of each design as well as a comparison between experiment and simulation jet velocities.

The designs parameterizations that we used in this analysis were limited to three or four parameters, which is a relatively low-dimensional parameter space. Future work will involve investigating higher-dimensional parameterizations and combining design parameterizations (e.g., a modified liner shape coupled with a silicone inclusion in the HE) to produce even more effective enhancement or suppression methods. A different design parameterization involving multiple detonators to produce more customizable detonation wave fronts, as described in Sterbentz et al.~\cite{Sterbentz2023}, may also be experimentally tested in the future. 

\section*{Acknowledgments}
This work was performed under the auspices of the U.S. Department of Energy by Lawrence Livermore National Laboratory under Contract DE-AC52-07NA27344 and was supported by the LLNL-LDRD Program under Project No. 21-SI-006. We would like to thank Justin Cassell, Alexander B.\ Kostinski, and William J.\ Schill for helpful discussions and guidance related to the simulation and theory presented in this paper. We also greatly appreciate the invaluable input on experiment design and part preparation from Robert V.\ Reeves and Michael D.\ Grapes. Lastly, we would like to acknowledge the significant amount of work performed by the High Explosives Applications Facility staff in support of this work, especially including Anthony B.\ Olson, Nathan W.\ Merriam, Jonathan C.\ Watkins, Caleb A.\ Glickman, Deanna M.\ Kahmke, Doug Lahowe, and James M.\ Clark. Document release number: LLNL-JRNL-862011.

\section*{Author Declarations}
\subsection*{Conflict of Interest}
The authors have no conflicts to disclose.

\section*{Data Availability}
The data that support the findings of this study are available from the corresponding author upon reasonable request.

\appendix
\section{Static FXR images and photonic Doppler velocimetry (PDV) data}\label{sec:PDV}
\begin{figure*}[!htp]
    \centering
    \includegraphics[width=10.0cm]{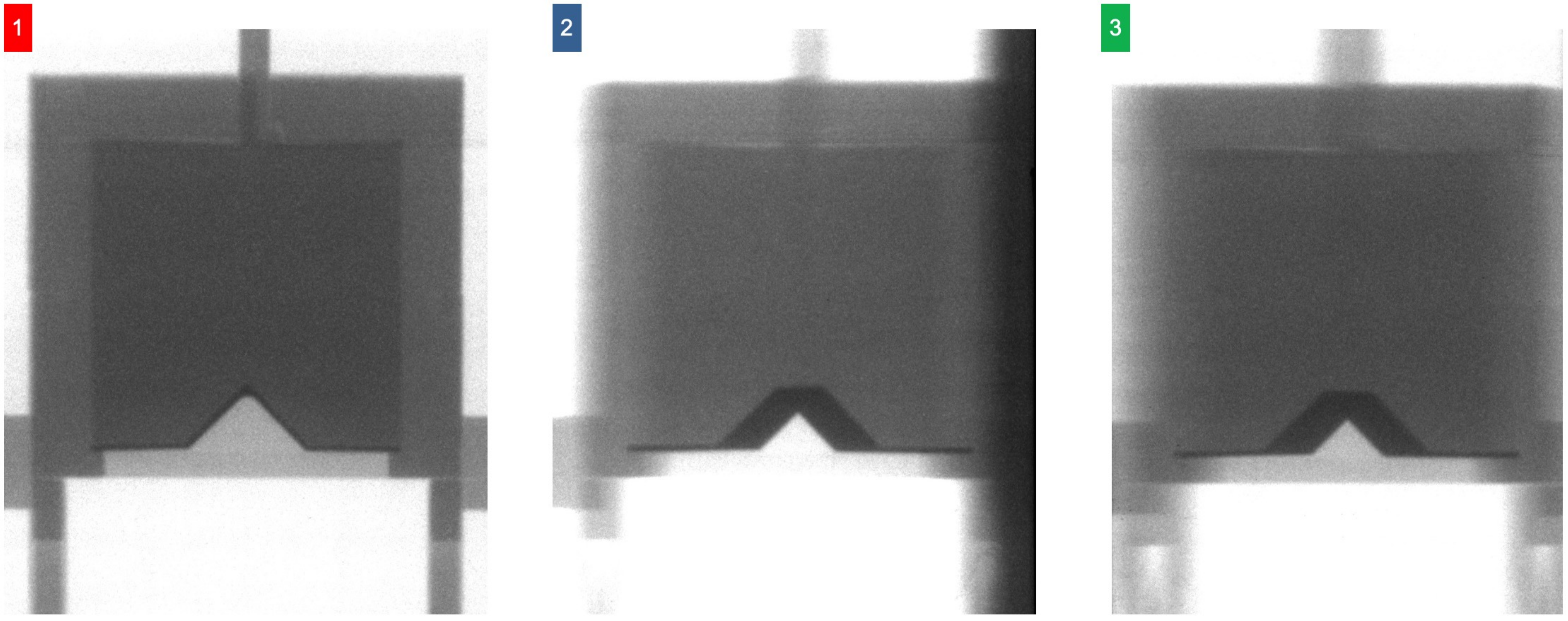}
    \qquad
    \centering
    
    \caption{Static FXR images of the baseline design prior to detonation.}
\label{fig:Fig_A1}
\end{figure*}

\begin{figure*}[!htp]
    \centering
    \includegraphics[width=10.0cm]{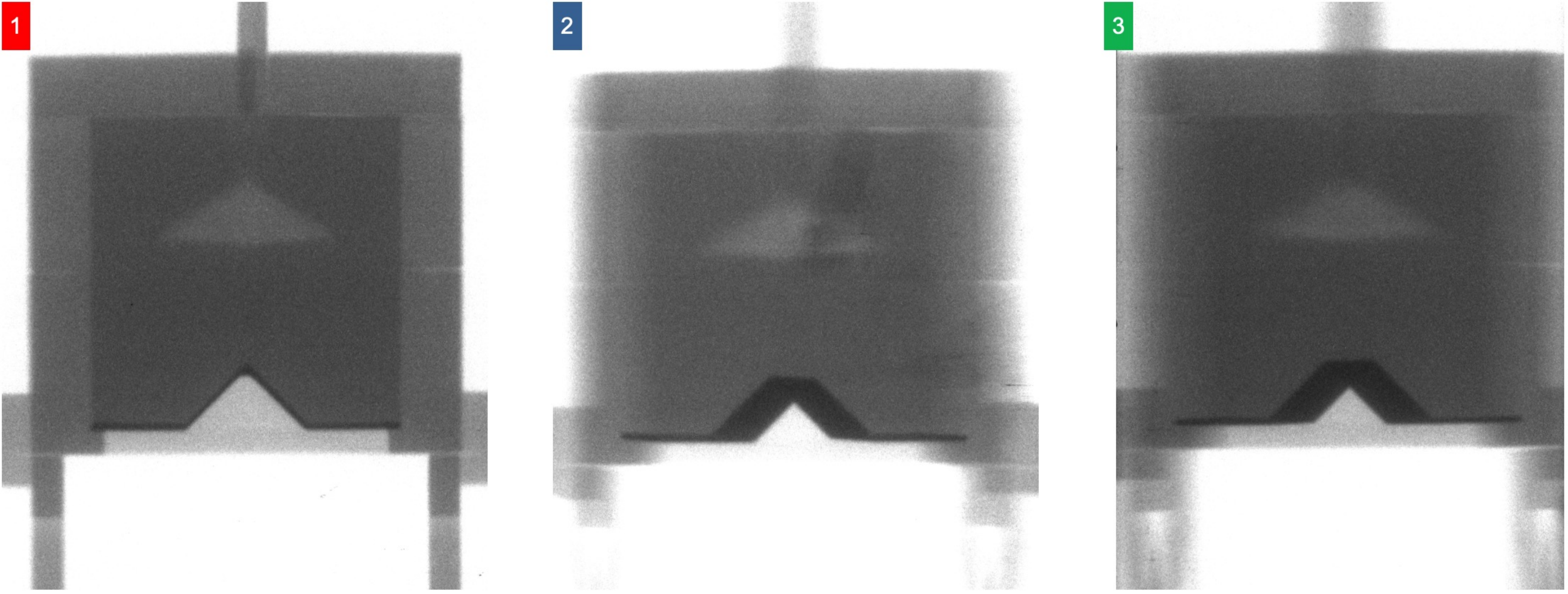}
    \qquad
    \centering
    
    \caption{Static FXR images of the jet enhancement design with a chevron silicone inclusion (Design 1) prior to detonation.}
\label{fig:Fig_A2}
\end{figure*}

\begin{figure*}[!htp]
    \centering
    \includegraphics[width=10.0cm]{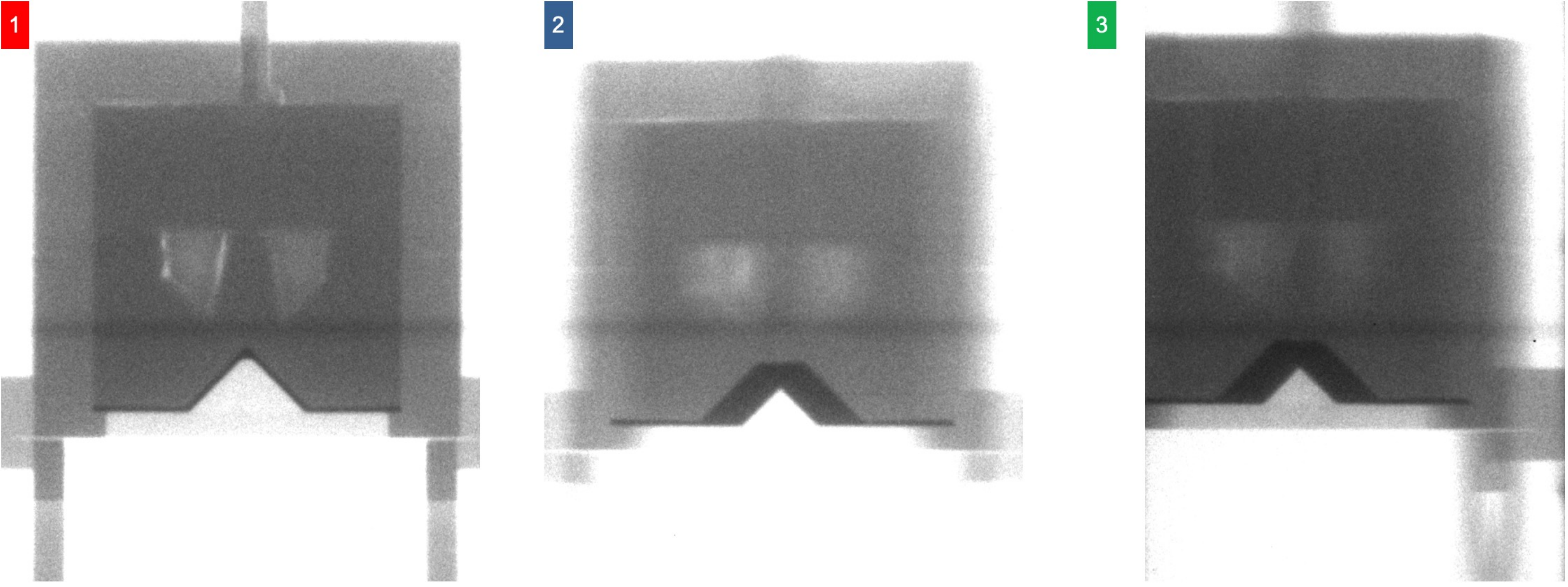}
    \qquad
    \centering
    
    \caption{Static FXR images of the jet suppression design with diffraction silicone inclusions (Design 2) prior to detonation.}
\label{fig:Fig_A3}
\end{figure*}

\begin{figure*}[!htp]
    \centering
    \includegraphics[width=10.0cm]{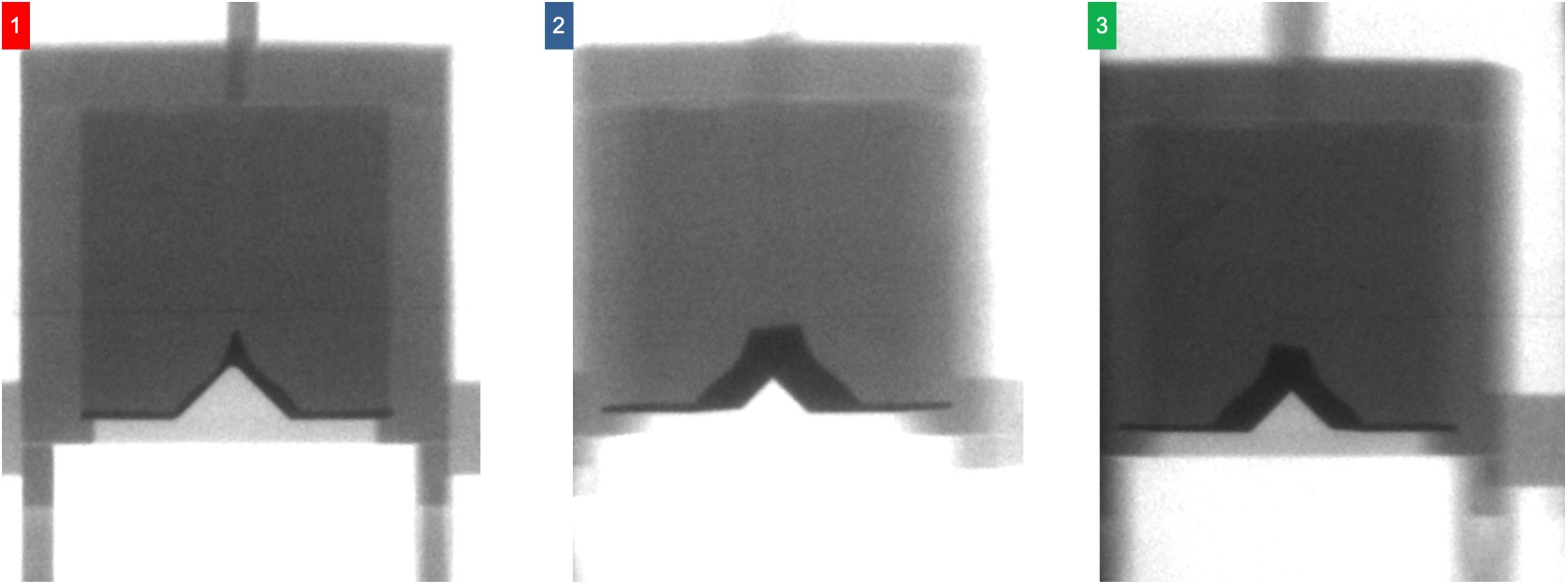}
    \qquad
    \centering
    
    \caption{Static FXR images of the jet enhancement liner modification design (Design 3) prior to detonation.}
\label{fig:Fig_A4}
\end{figure*}

\begin{figure*}[!htp]
    \centering
    \includegraphics[width=10.0cm]{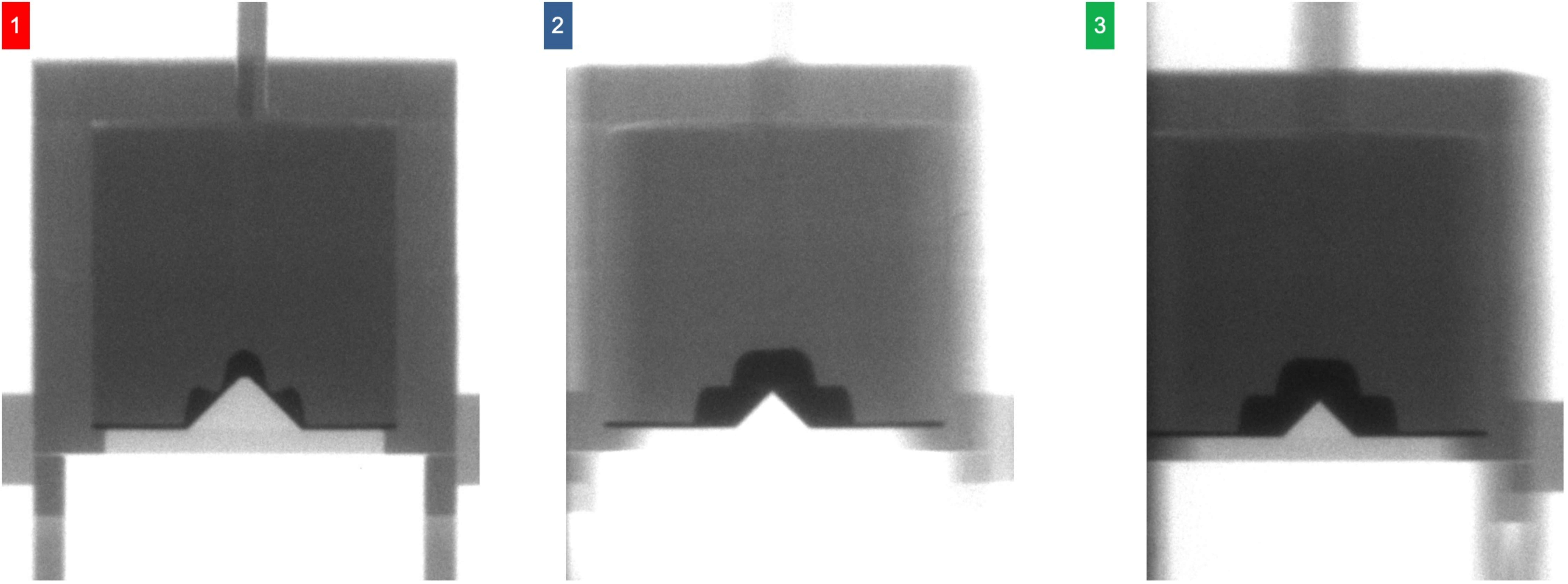}
    \qquad
    \centering
    
    \caption{Static FXR images of the jet suppression liner modification design (Design 4) prior to detonation.}
\label{fig:Fig_A5}
\end{figure*}

The experiment diagnostics include both FXR radiography images taken at three times and twelve PDV probes that measure the velocity at locations near the apex of the liner (i.e., where jetting occurs) as well as along the flat shoulder of the liner. Three static FXR images using the same radiography diagnostic setup shown in Figure \ref{fig:exp_layout}(a) (i.e., at angles of 0$^\mathrm{o}$ and $\pm$8$^\mathrm{o}$) are also taken prior to detonation for the baseline case and each of the designs. These static FXR images are shown in Figures \ref{fig:Fig_A1}--\ref{fig:Fig_A5}. 

\begin{figure}[!htp]    
    \centering
    \subfloat[]{{\includegraphics[width=7.5cm]{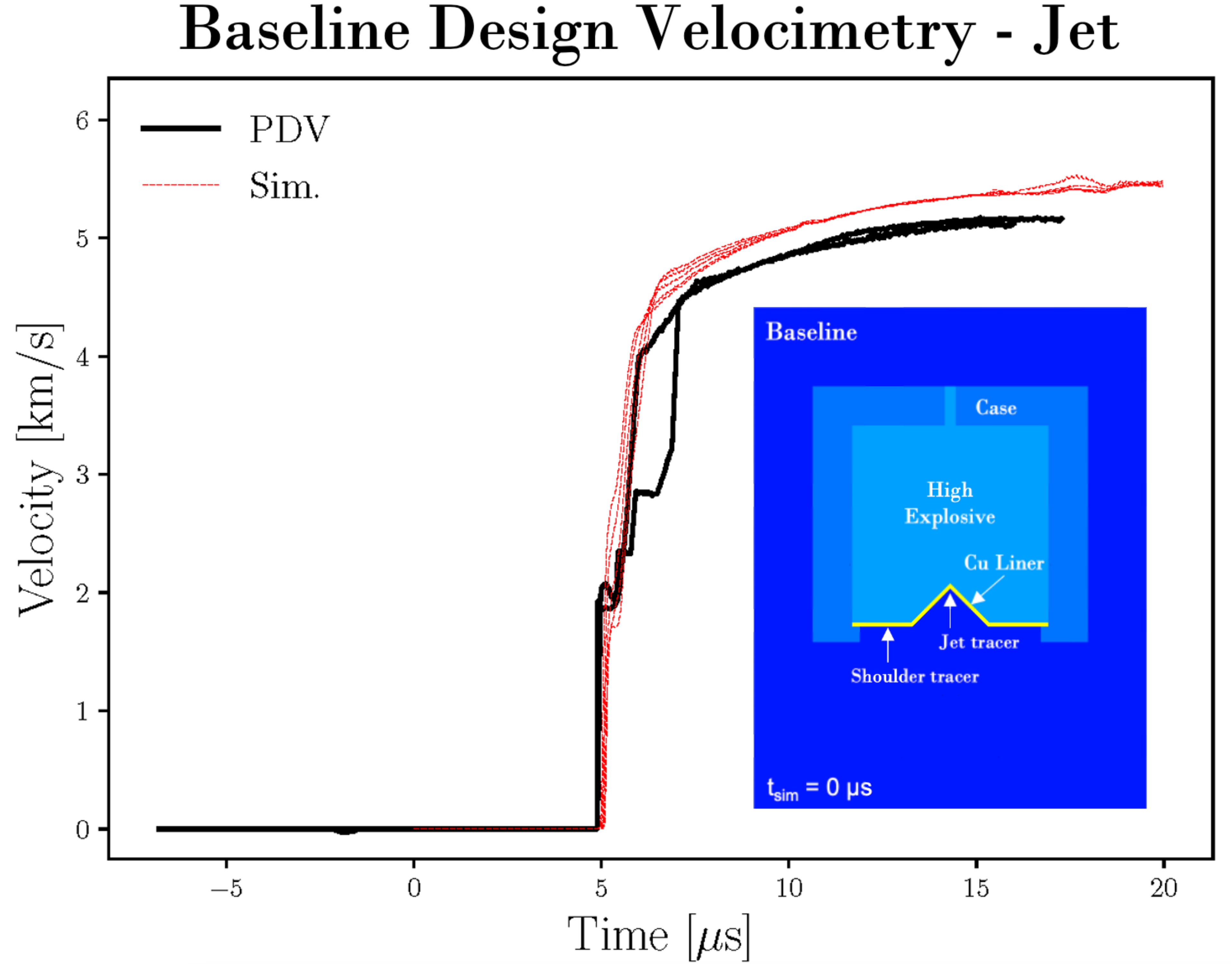} }}
    \qquad
    \centering
    \subfloat[]{{\includegraphics[width=7.5cm]{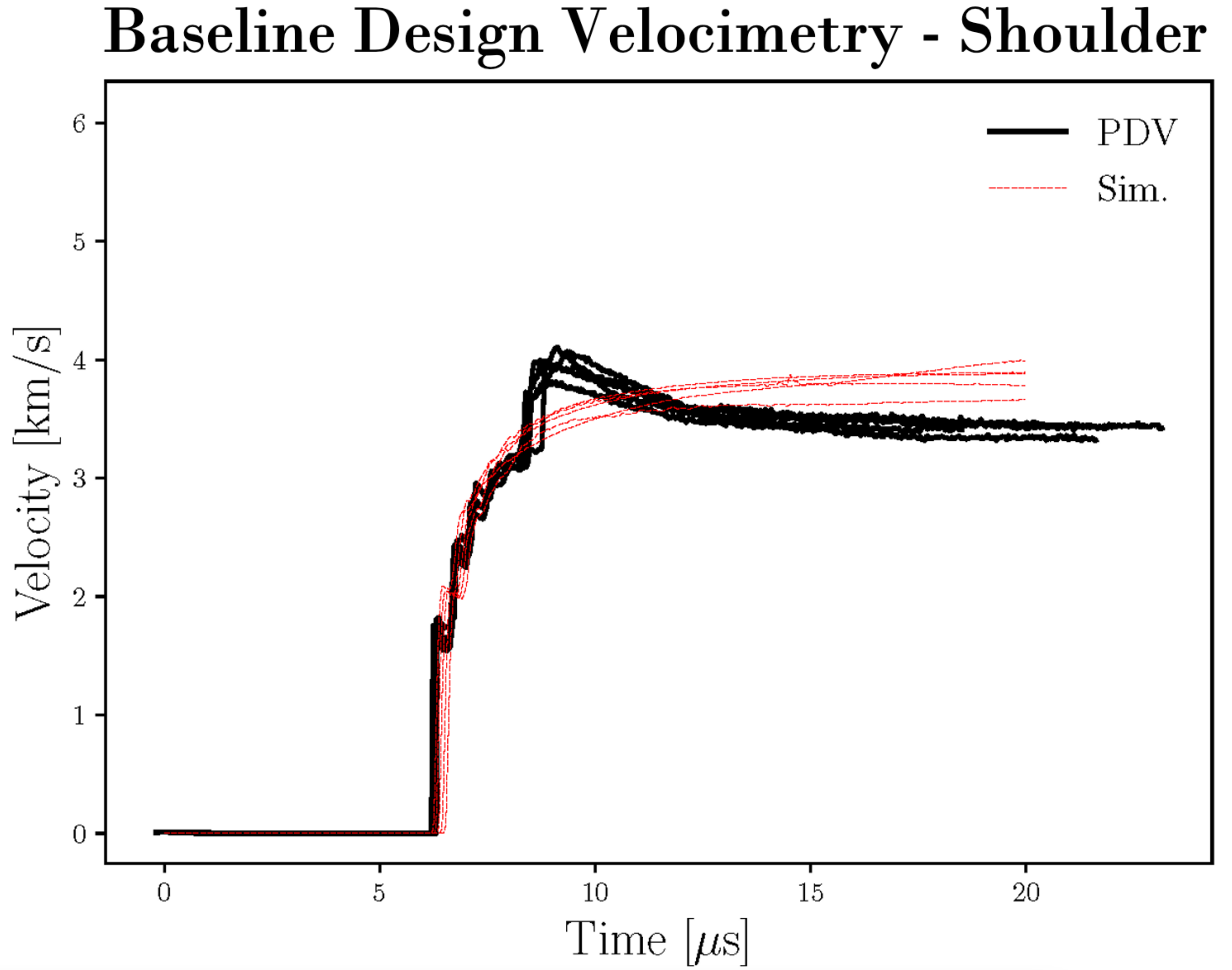} }}
    \qquad
    \centering
    
    \caption{The PDV velocity data versus time compared to the simulation tracer data from several locations on the liner. The plots in this figure correspond to the baseline case for: (a) PDV probes and tracers near jet; (b) PDV probes and tracers near liner shoulder (flat region of liner).}
\label{fig:PDV_baseline}
\end{figure}

\begin{figure}[!htp]
    \centering
    \subfloat[]{{\includegraphics[width=7.5cm]{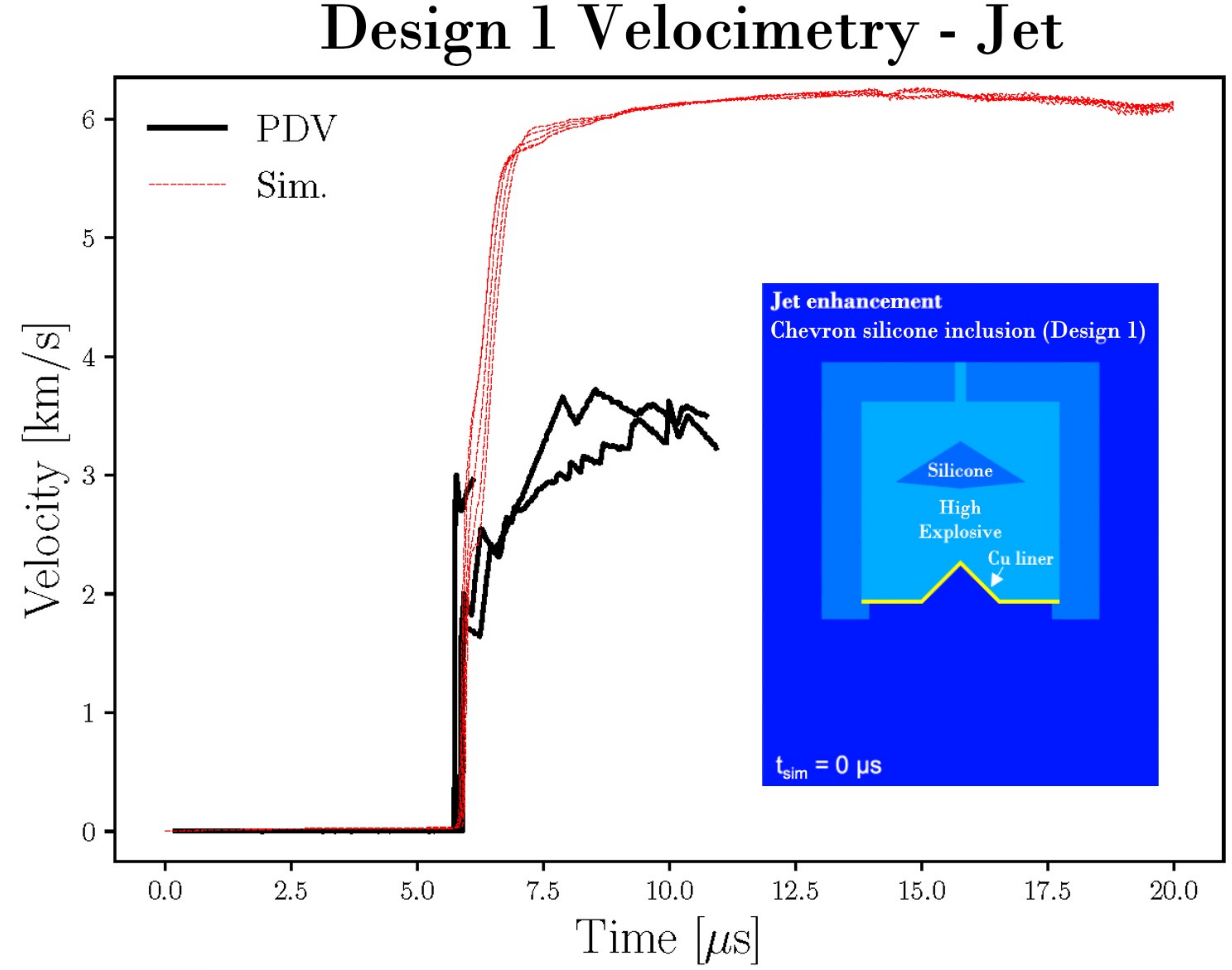} }}
    \qquad
    \centering
    \subfloat[]{{\includegraphics[width=7.5cm]{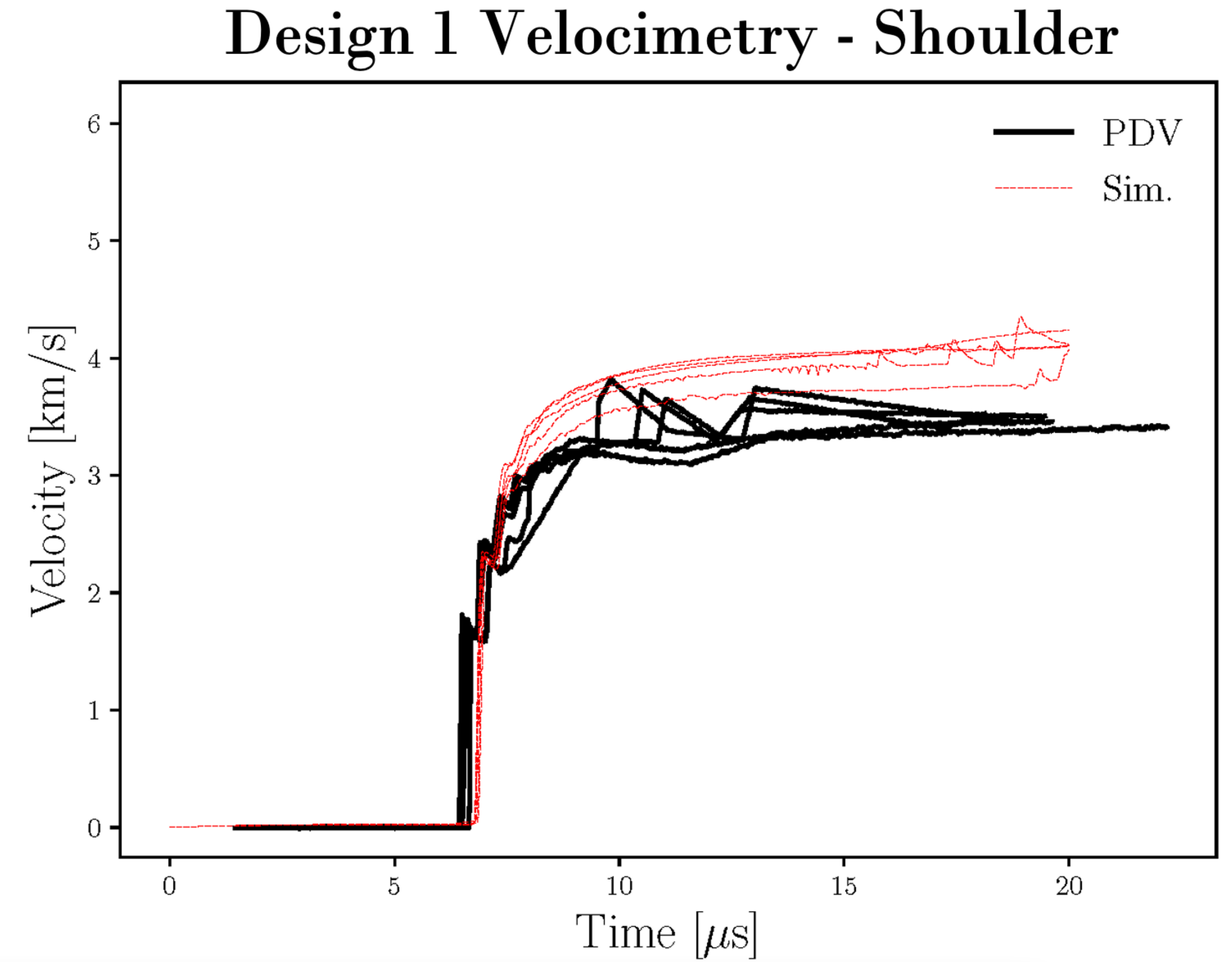} }}
    \qquad
    \centering
    
    \caption{The PDV velocity data versus time compared to the simulation tracer data from several locations on the liner. The plots in this figure correspond to Design 1 (chevron silicone inclusion, jet enhancement) for: (a) PDV probes and tracers near jet; (b) PDV probes and tracers near liner shoulder (flat region of liner).}
\label{fig:PDV_chevron}
\end{figure}

\begin{figure}[!htp]
    \centering
    \subfloat[]{{\includegraphics[width=7.5cm]{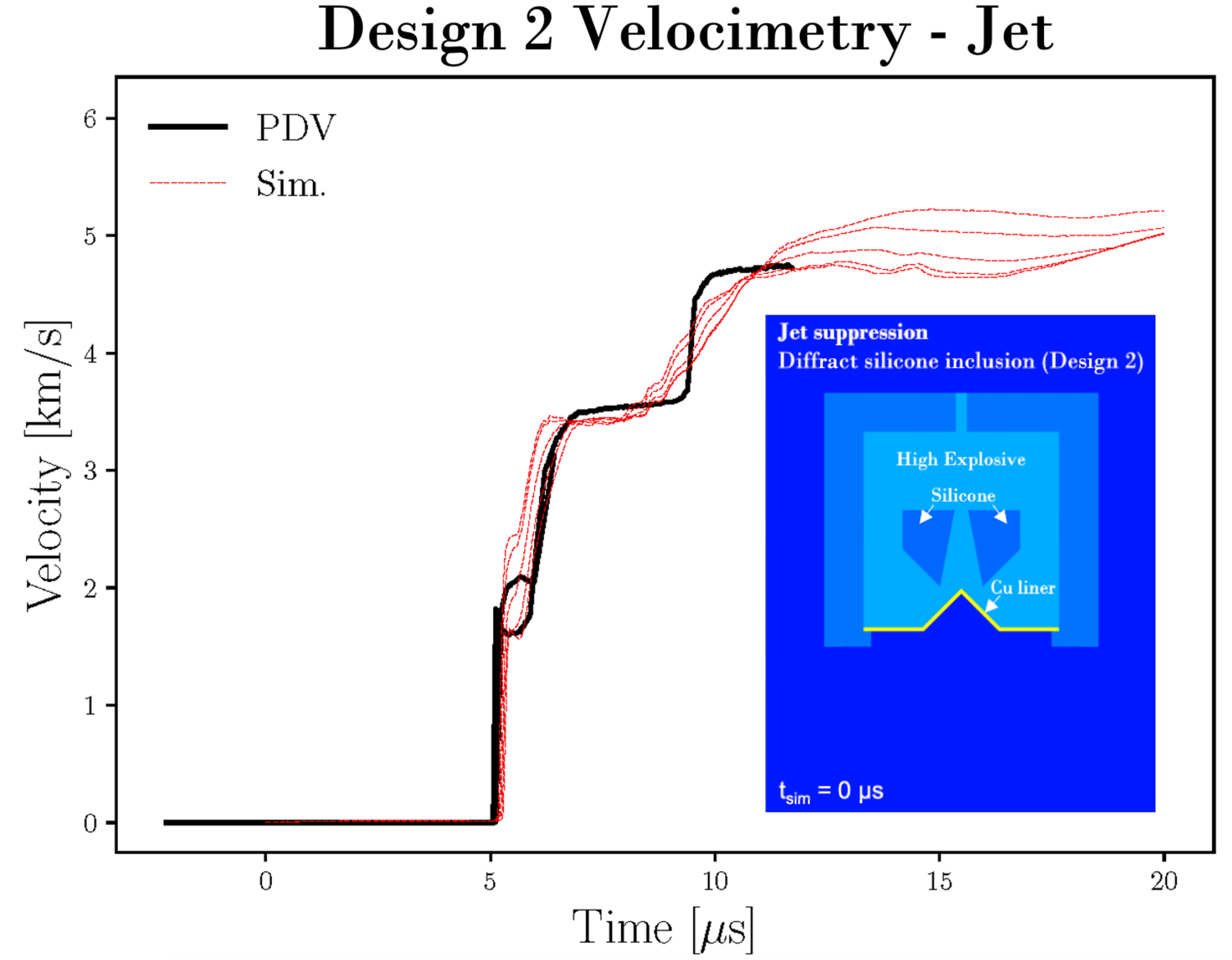} }}
    \qquad
    \centering
    \subfloat[]{{\includegraphics[width=7.5cm]{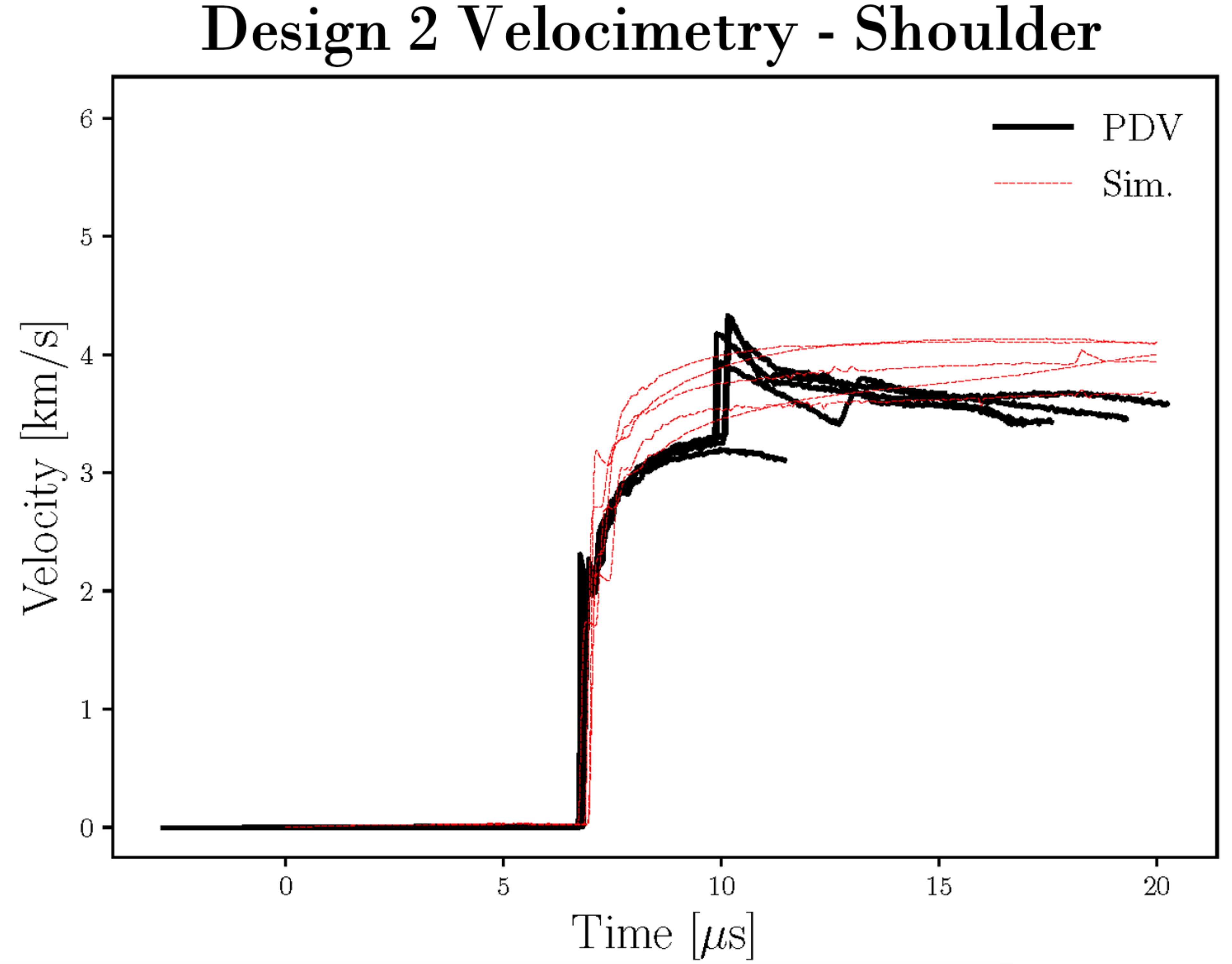} }}
    \qquad
    \centering
    
    \caption{The PDV velocity data versus time compared to the simulation tracer data from several locations on the liner. The plots in this figure correspond to Design 2 (diffraction silicone inclusions, jet enhancement) for: (a) PDV probes and tracers near jet; (b) PDV probes and tracers near liner shoulder (flat region of liner).}
\label{fig:PDV_diffract}
\end{figure}

\begin{figure}[!htp]
    \centering
    \subfloat[]{{\includegraphics[width=7.5cm]{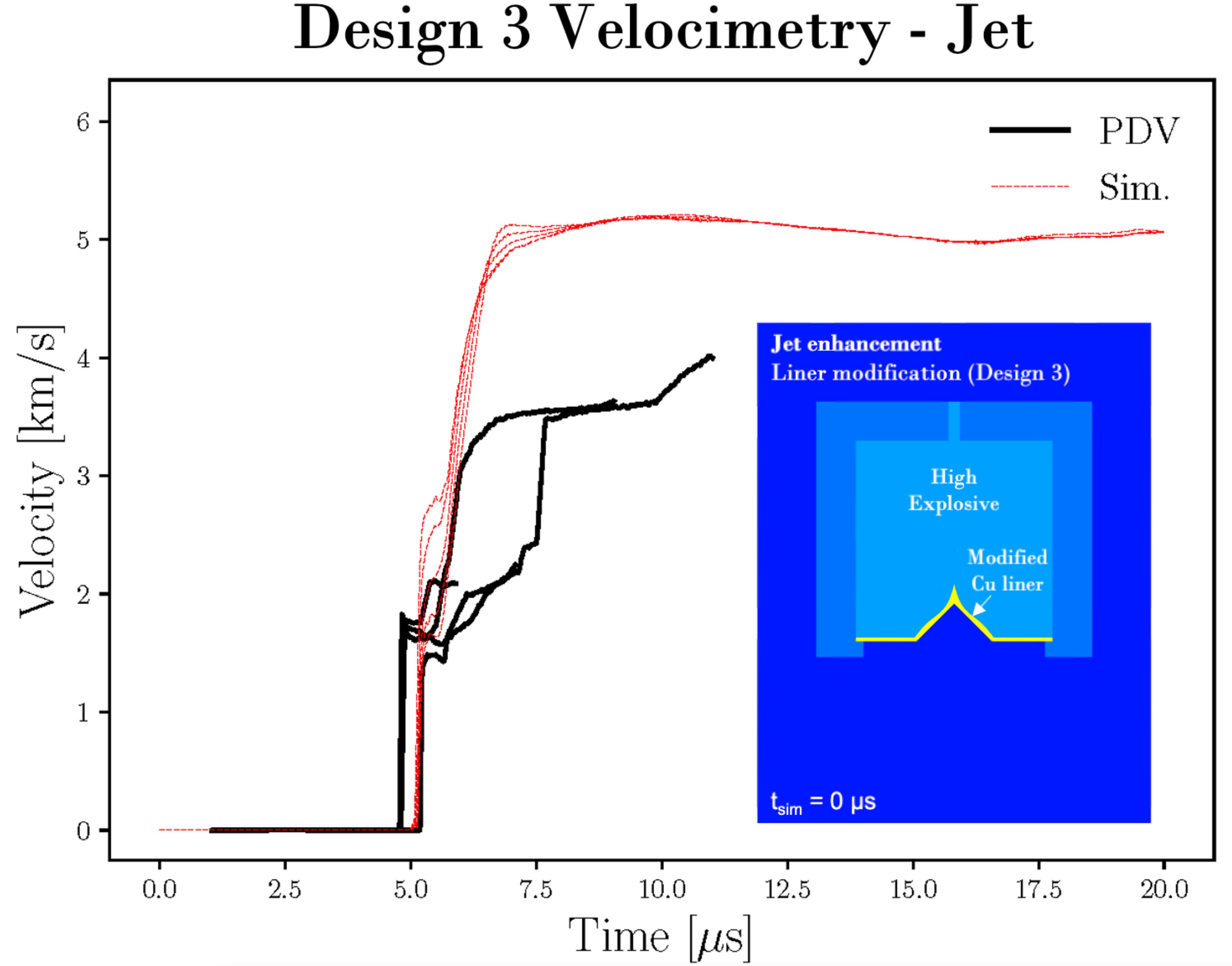} }}
    \qquad
    \centering
    \subfloat[]{{\includegraphics[width=7.5cm]{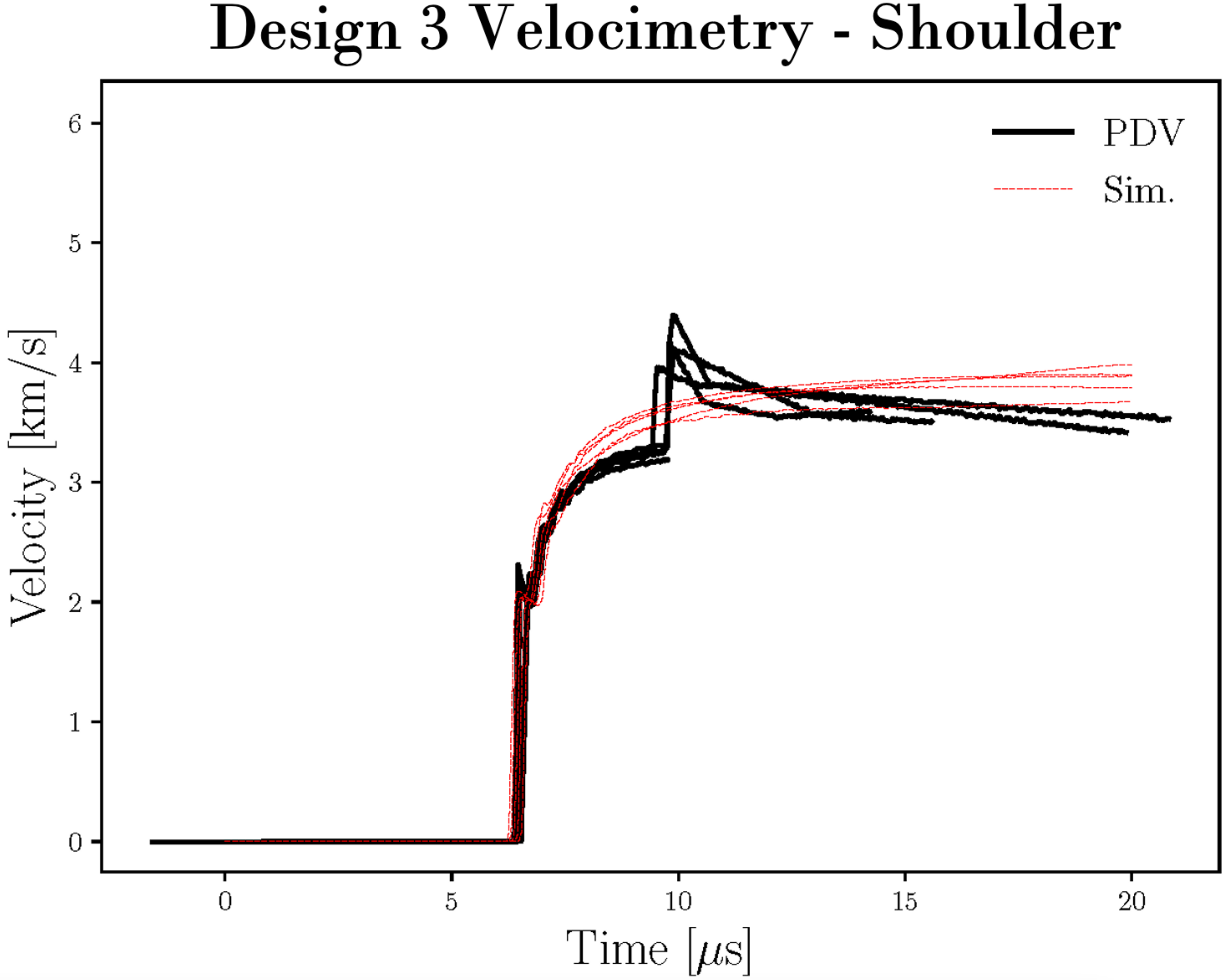} }}
    \qquad
    \centering
    
    \caption{The PDV velocity data versus time compared to the simulation tracer data from several locations on the liner. The plots in this figure correspond to Design 3 (liner modification, jet enhancement) for: (a) PDV probes and tracers near jet; (b) PDV probes and tracers near liner shoulder (flat region of liner).}
\label{fig:PDV_liner_max}
\end{figure}

\begin{figure}[!htp]
    \centering
    \subfloat[]{{\includegraphics[width=7.5cm]{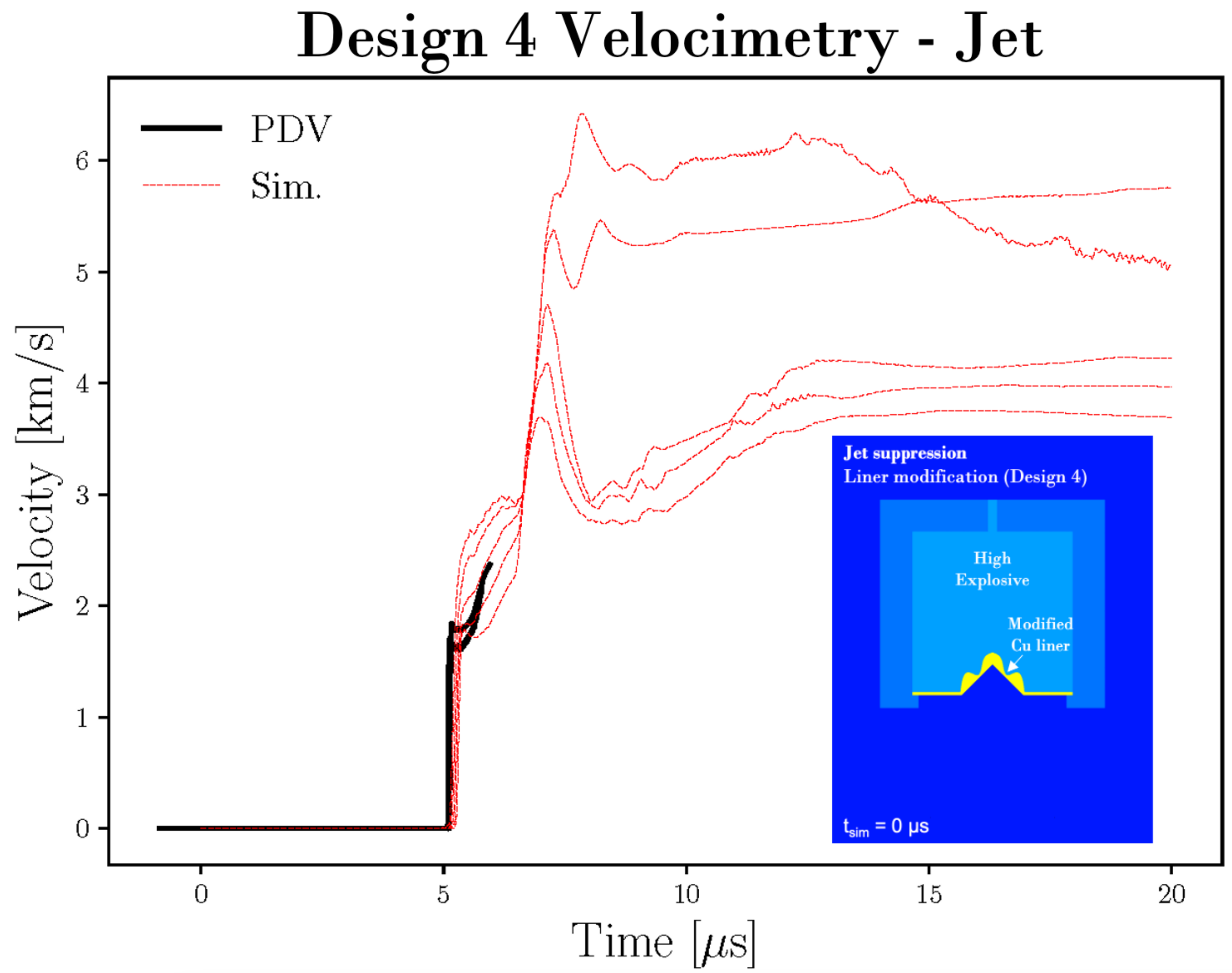} }}
    \qquad
    \centering
    \subfloat[]{{\includegraphics[width=7.5cm]{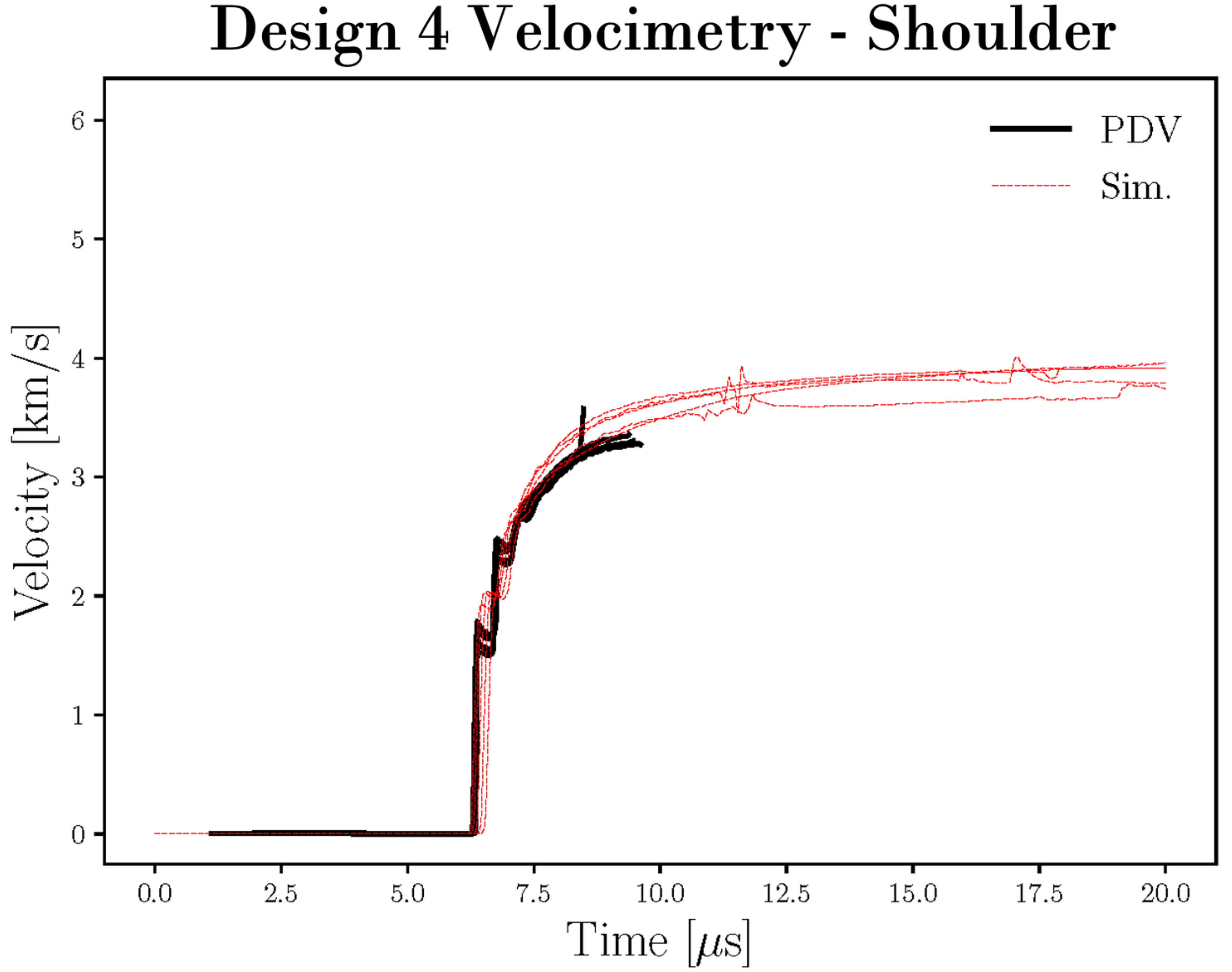} }}
    \qquad
    \centering
    
    \caption{The PDV velocity data versus time compared to the simulation tracer data from several locations on the liner. The plots in this figure correspond to Design 4 (liner modification, jet suppression) for: (a) PDV probes and tracers near jet; (b) PDV probes and tracers near liner shoulder (flat region of liner).}
\label{fig:PDV_liner_min}
\end{figure}

The PDV velocity measurements can be compared to tracer data from simulations. The location of the PDV measurement point along the liner does not always correspond to a single simulation tracer location and likely varies as the liner mass is formed into a jet. For this reason, we have plotted several PDV and tracer data curves, which provides a better comparison between the two than simply using the data from a single simulation tracer. Figure \ref{fig:PDV_baseline} shows a comparison between PDV measurements and simulation tracer data over time for the baseline case. This includes the data near the jet tip in Figure \ref{fig:PDV_baseline}(a) and at the flat liner shoulder in Figure \ref{fig:PDV_baseline}(b). Figure \ref{fig:PDV_chevron} and \ref{fig:PDV_diffract} show the PDV and simulation tracer data for Design 1 (enhancement) and Design 2 (suppression), respectively. Lastly, Figure \ref{fig:PDV_liner_max} and \ref{fig:PDV_liner_min} provide the PDV and simulation tracer data for Design 3 (enhancement) and Design 4 (suppression).

For the experiments in which a narrow high-velocity jet was formed, the PDV probes tend to not capture the maximum jet velocity. This is why there appears to be a large discrepancy between the experiment PDV measurements and the simulation tracer data in Figures \ref{fig:PDV_chevron}(a) and \ref{fig:PDV_liner_max}(a). This may be due to small defects in the production of these experiments causing the jet tip to drift or ``tilt" to one side and the PDV probe to miss the jet tip. Additionally, for the liner suppression case (Design 4) in Figure \ref{fig:PDV_liner_min}, there is no distinct jet that forms and the PDV probes do not capture any data after approximately 6~$\mathrm{\mu}$s after detonation in the experiment. Instead, a high-velocity low-density cloud of liner ``dust" is formed that is captured by the simulation tracers, which accounts for the discrepancy in Figure \ref{fig:PDV_liner_min}(a). However, for the other cases, there is reasonable agreement between the PDV probe data and the simulation tracer data, particularly at the liner shoulder.

\bibliographystyle{elsarticle-num}

\end{document}